\documentclass[a4paper,12pt]{article}
\usepackage{lmodern}
\usepackage[left=2.5cm,right=2.5cm,top=3cm,bottom=3cm]{geometry}
\usepackage[T1]{fontenc}
\usepackage[utf8]{inputenc}
\usepackage[english]{babel}
\usepackage{amsmath,amssymb,amsthm,mathtools,bm}
\usepackage{graphicx}
\usepackage[font=small,labelfont=bf]{caption}
\usepackage{tabularx,booktabs}
\usepackage{adjustbox}
\usepackage{titling}
\usepackage{authblk}
\usepackage{titlesec}

\numberwithin{equation}{section}
\newtheorem{theorem}{Theorem}
\newtheorem{proposition}{Proposition}
\newtheorem{remark}{Remark}

\usepackage[dvipsnames]{xcolor}
\usepackage{tikz-cd}
\usepackage{tikz}
\usetikzlibrary{positioning}
\usetikzlibrary{calc}

\definecolor{mylightred}{RGB}{211,79,73}
\definecolor{mydarkred}{RGB}{199,44,38}
\definecolor{mylightgreen}{RGB}{78,153,67}
\definecolor{mydarkgreen}{RGB}{43,129,33}
\definecolor{mylightpurple}{RGB}{150,107,178}
\definecolor{mydarkpurple}{RGB}{126,78,160}
\definecolor{mylightblue}{RGB}{49,101,205}
\definecolor{mydarkblue}{RGB}{20,92,205}

\usepackage[hidelinks]{hyperref}

\definecolor{magmaDark}{HTML}{000004}
\definecolor{magmaViolet}{HTML}{5c1a87}
\definecolor{magmaMagenta}{HTML}{721f81}
\definecolor{magmaRose}{HTML}{cd4071}
\definecolor{magmaOrange}{HTML}{fd9668}
\definecolor{magmaPeach}{HTML}{feca8d}

\hypersetup{
pdfstartview=FitH,
colorlinks=true,
linkcolor=magmaViolet,
urlcolor=magmaOrange,
citecolor=magmaRose}

\tikzset{
  juliadot/.style args={#1,#2}{shape=circle,line width=0.03ex,minimum width=0.4ex,fill=#1,draw=#2}
}

\newcommand\julialetter[1]{{\strut\fontfamily{cmss}\bfseries\selectfont{#1}}}

\DeclareRobustCommand\julia{\begin{tikzpicture}[baseline=0mm, every node/.style={inner sep=0mm, outer sep=0mm}]
\node[anchor=base]        (j) at (0,0) {\julialetter{\j}};
\node[anchor=base, right=0ex of j] (u) {\julialetter{u}};
\node[anchor=base, right=0ex of u] (l) {\julialetter{l}};
\node[anchor=base, right=0ex of l] (i) {\julialetter{\i}};
\node[anchor=base, right=0ex of i] (a) {\julialetter{a}};
\path let \p1 = (j) in node[juliadot={mylightblue,mydarkblue}] (bluedot) at (\x1+0.02ex,1.4ex) {};
\path let \p1 = (i) in node[juliadot={mylightred,mydarkred}] (reddot) at (\x1,1.4ex) {};
\path let \p1 = (reddot) in node[juliadot={mylightpurple,mydarkpurple}] (purpledot) at (\x1+0.5ex,\y1) {};
\path let \p1 = (reddot) in node[juliadot={mylightgreen,mydarkgreen}] (greendot) at (\x1+0.25ex,\y1+0.42ex) {};
\end{tikzpicture}}

\renewcommand{\phi}{\varphi}
\DeclareMathOperator{\tr}{\mathrm{tr}}
\DeclareMathOperator{\re}{\mathrm{Re}}
\DeclareMathOperator{\im}{\mathrm{Im}}

\pretitle{\centering\large\bfseries}
\posttitle{\par\vspace{0.6em}}

\preauthor{\centering\normalsize}
\postauthor{\par\vspace{0.4em}}

\predate{\centering\small}
\postdate{\par}

\titleformat*{\section}{\large\bfseries}
\titleformat*{\subsection}{\normalsize\bfseries}

\newcommand{\keywords}[1]{  \par\smallskip\noindent{\small\textbf{Keywords:} #1\par}}

\addto\captionsenglish{}

\begin{document}

\title{Fate of Arnold tongues and canards in M\"{o}bius systems with noise}
\author[1,2,3]{\normalsize Artem Alexandrov\thanks{alexandrov.aa@mi-ras.ru}}

\affil[1]{Steklov Mathematical Institute of Russian Academy of Sciences, Moscow, 119333, Russia}

\affil[2]{Phystech School of Applied Mathematics and Computer Science, Moscow Institute of Physics and Technology, Dolgoprudny, 141700, Russia}

\affil[3]{Laboratory of Complex Networks, Center for Neurophysics and Neuromorphic Technologies, Moscow, Russia}

\date{}

\maketitle

\begin{abstract}
We consider a class of noisy dynamical systems on the two-dimensional torus whose phase evolution is governed by M\"{o}bius transformations. In the deterministic limit, these systems favor phase-locking that is described by the band-gap structure of an associated Schr\"{o}dinger equation with a periodic potential. We use the Wentzel-Kramers-Brillouin method to characterize this structure in the slow-fast limit. Then we show that large-deviation theory describes fluctuations of the finite-time rotation number, whereas weak-noise escape actions determine the exponential scale of rare phase slips. These estimates provide quantitative criteria for metastable phase-locking and melting of Arnold tongues over a prescribed observation time. We derive sufficient conditions for stochastic trajectories to follow prescribed canard segments. With controlled entrance errors, exponential amplification of fluctuations along repelling branches leads to an exponentially small sufficient noise scale for canard persistence in the slow-forcing limit.

\end{abstract}

\newpage
\tableofcontents
\newpage

\section{Introduction}

Singular perturbation theory appears in different forms in nonlinear dynamics, quantum mechanics, and probability theory. For example, geometric slow-fast methods describe motion near slow manifolds~\cite{fenichel1979geometric,kuehn2015timescales,wechselberger2020nonstandard,kristiansen2023review}, the Wentzel-Kramers-Brillouin (WKB) method describes semiclassical propagation and tunneling~\cite{delabaere1997semiclassical,iwaki2025lectures}, and the Freidlin-Wentzell (FW) theory describes exponentially rare fluctuations under weak noise~\cite{freidlin2012random,dembo2010deviations,grafke2019rare}. These approaches use different languages, but in each setting exponentially small effects can determine the observable dynamics. Here we bring the three approaches together to control deterministic canard selection, describe noisy tracking, and explain metastable phase-locking in periodically driven M\"{o}bius systems.

The M\"{o}bius systems describe the evolution of the phase variable under the action of an external drive. The interaction between the phase variable and the external force is captured by the phase-locking phenomenon, which can be characterized by the Poincar\'{e} rotation number. In particular, phase-locking plateaus correspond to an integer rotation number for M\"{o}bius systems. From the viewpoint of non-equilibrium statistical mechanics, phase winding is an integrated transport current, and the rotation number is its asymptotic rate in turns per forcing period~\cite{tsobgninyawo2016current}. Long time large deviations describe fluctuations of this current at fixed noise~\cite{touchette2009deviations,touchette2018markov,burenev2025deviations}. This interpretation places canard persistence in the context of rare transport events and describes noise-induced rounding (we call it \emph{melting}) of locking plateaus.

Canards are trajectories that follow a repelling branch of the slow manifold for an $O(1)$ interval of slow time. Pioneering work by Beno\^{i}t, Callot and Dieners uncovered these trajectories in the van der Pol oscillator using nonstandard analysis~\cite{benoit1981chasse,diener1984unchained}. Canard explosions~\cite{krupa2001relaxation} describe the rapid growth from small oscillations to relaxation oscillations over an exponentially narrow parameter interval. Subsequent works developed the theory in three-dimensional systems~\cite{benoit1983systemes}, examined the role of curvature~\cite{desroches2011curvature}, and identified canards in neuronal and chemical models~\cite{moehlis2006hodgkinhuxley,brons1991belousov}. Geometric singular perturbation theory and blow-up methods provide the foundations for analyzing fold points and their relation to canards~\cite{krupa2001fold}, with more recent extensions to reaction-diffusion equations and neural fields~\cite{avitabile2020spatiotemporal}. The comprehensive treatment of canards is given in~\cite{dumortier1996cycles,demaesschalck2021canard}. The stability of a canard depends on the accumulated contraction and expansion, while exponentially small changes in initial conditions or parameters can determine the departure from a repelling branch.

Especially relevant here are Callot's studies of Hermite-type and Schr\"{o}dinger equations~\cite{callot1981bifurcations,callot1985visibles}, together with his analysis of complex slow-fast fields and associated Riccati equations~\cite{callot1993champs}. In addition to these works, Beno\^{i}t, El Hamidi, and Fruchard used combined asymptotic expansions to relate exponentially close canard values to energy-level splitting in a symmetric double-well potential~\cite{benoit2002combined}. Subsequently, Pavis d'Escurac studied the Borel summation of formal canard-parameter expansions and the exponentially small differences between sectorial solutions~\cite{pavisdescurac2018singularities,pavisdescurac2018borel}. These works establish direct connections between canards, spectral splitting, and asymptotic summation.

The geometry that we considered here makes canard study particularly accessible. A periodically driven phase defines an autonomous deterministic system on the two-torus, with one fast variable $\phi$ and one slow variable $\tau$, where $\dot\tau=\omega$, and forcing frequency is $0<\omega\ll 1$. Therefore, we have the slow-fast system and $\omega\to 0$ limit is singular. The canard cycles in this setting were related to staircase structures by Guckenheimer and Ilyashenko and then more generally studied by Schurov~\cite{guckenheimer2001staircase,schurov2010torus}. We focus on the M\"{o}bius subclass, defined precisely below. Its Poincar\'{e} (return) map is fractional-linear map and can be analyzed through a two-dimensional linear system~\cite{marvel2009mobius}.

This subclass arises in the leading order in phase reduction near a supercritical Hopf bifurcation under weak additive forcing~\cite{nakao2016synchronisation}. For example, in the description of self-sustained oscillators, such as the Ermentrout-Kopell $\theta$-neuron normal form for type-I excitability~\cite{ashwin2016hopf,ermentrout1986bursting}. This connection motivates our choice of M\"{o}bius subclass, while the transferring of quantitative predictions for exponentially narrow canard windows to general non-linear oscillators requires control of how the terms omitted by phase reduction affect their location and width.

Random perturbations introduce a second singular limit. The sample-path estimates for stochastic slow-fast systems were investigated by Berglund and Gentz in~\cite{berglund2003geometric,berglund2006noise}. Then, Sowers studied stochastic canard selection~\cite{sowers2008canards}, while Berglund, Gentz, and Kuehn analyzed early departures near folded nodes with two slow variables~\cite{berglund2012hunting}. Escape through unstable periodic orbits and phase-slip statistics provide a complementary description of rare loss of synchrony~\cite{berglund2014passage,berglund2016phaseslips}. These results motivate distinguishing departure from an individual canard excursion from the loss of phase-locking over many cycles.

The M\"{o}bius structure provides the connection to spectral theory. Riccati linearization and scalar reduction in suitable charts relate deterministic dynamics to a Schr\"{o}dinger-type equation with periodic potential, while the Floquet monodromy connects phase-locking domains to spectral gaps, as discussed in~\cite{alexandrov2026phaselock}. Rotation numbers and cocycles provide a broader spectral context through the work of Johnson and Moser~\cite{johnson1982rotation} and Avila's contributions to gap opening and quasiperiodic operators~\cite{avila2012gaps,avila2015global}. Here, we consider periodic forcing, and WKB resolves the exponentially small scales, which are needed for canard selection.

Exact WKB and its resurgent formulation recover exponentially small contributions through Borel summation and connection formulas~\cite{delabaere1997semiclassical,dunne2017mathieu,bucciotti2023anharmonic}, see Iwaki's Les Houches lectures that provide a modern introduction~\cite{iwaki2025lectures}. Their relation to semiclassical saddle contributions and quantization is explicit in studies of multi-well and periodic quantum potentials~\cite{sueishi2020quantization,sueishi2021anomaly,misumi2025degenerate,misumi2026nondegenerate}. Rigorous results that establish Borel resummed Riccati and WKB solutions and describe the geometry of their Borel singularities are given in~\cite{nikolaev2023riccati,nikolaev2023wkb,nikolaev2024geometry}. The complementary view on turning-point connections and Bohr-Sommerfeld quantization is based on the use of slow-fast geometry and blow-up methods~\cite{kristiansen2024turning,kristiansen2025eigenvalue}.

The deterministic part of our analysis translates spectral information into trajectory geometry. In the slow-fast limit, WKB actions and connection formulas determine the relative weights of growing and decaying solutions, which define the available repelling branch excursion. The WKB estimates resolve an exponentially narrow canard-supporting layer inside a spectral gap. At the relevant simple band edges, the limiting periodic trajectories have unit multipliers of the phase return map, so that the accumulated contraction and expansion balance over a period. In the analyzed canard regime, these trajectories follow attracting and repelling slow branches and form balanced canards.

The stochastic part combines this selection with linearized fluctuation estimates and FW actions. Linearized fluctuations quantify noise amplification along the selected excursion, while weak-noise actions control rare phase slips. This weak-noise limit is distinct from the long-time current large deviations mentioned above. At fixed forcing frequency and relative noise-channel strengths, the rate of a specified phase-slip event in the metastable regime has a leading exponential factor $\exp[-S^{\mathrm{FW}}/D]$, where $D$ is the overall noise intensity and $S^{\mathrm{FW}}$ is the minimum action for that event. Unlike the deterministic WKB action, this action depends on the noise geometry.

These deterministic and stochastic studies complement each other. They give conditions for tracking a selected canard excursion and for metastable locking. We illustrate our findings with several examples. We use the Adler equation~\cite{adler1946locking} as the simplest possible phase-locking benchmark. More substantive examples are the periodically forced $\theta$-neuron~\cite{ermentrout1986bursting}, the overdamped Josephson junction (RSJ-model)~\cite{mccumber1968rsj,stewart1968rcsj}, and a constructed M\"{o}bius model (we call it G2-model and will specify further). These examples emphasize that the described phase dynamics can arise in different scenarios. Building on~\cite{alexandrov2026duck}, we connect deterministic canard selection with two distinct questions of stochastic persistence. We combine WKB connection data with fluctuation estimates to quantify finite-time tracking of selected canard excursions, and use event-dependent FW actions to characterize rare losses of phase-locking. This gives separate conditions for observing an individual canard excursion and for maintaining metastable locking over many forcing periods. Each section provides the essential facts and key results. To guide the reader, we provide the roadmap,

\begin{table}[h!]
    \centering
    \small
    \setlength{\tabcolsep}{5pt}
    \renewcommand{\arraystretch}{1.15}
    \begin{tabularx}{\linewidth}{@{}l >{\raggedright\arraybackslash}p{0.32\linewidth} >{\raggedright\arraybackslash}X@{}}
        \toprule
        Section & Question & Focus \\
        \midrule
        \ref{sec:Mobius-systems} & How is phase-locking can be described with noise? & M\"{o}bius structure, rotation number, phase slips, and metastable locking \\
        \addlinespace
        \ref{sec:WKB-deterministic-skeleton} & How are deterministic tongues constructed? & Spectral gaps, WKB actions, and monodromy \\
        \addlinespace
        \ref{sec:LDP-exponential-scale} & What sets the cost of rare fluctuations? & Event-dependent Freidlin-Wentzell actions and comparison of exponential scales \\
        \addlinespace
        \ref{sec:canard-persistence} & When does a canard persist under noise? & Entrance accuracy, noise amplification, and finite-time persistence \\
        \bottomrule
    \end{tabularx}
\end{table}

\section{M\"{o}bius systems and their stochastic extension}
\label{sec:Mobius-systems}

To start with, we recall the deterministic M\"{o}bius structure and its relation to phase-locking and to the Schr\"{o}dinger equation with periodic potential. Then, we introduce noise that preserves this structure pathwise, establish the existence of the rotation number, and describe its finite-time fluctuations. The weak-noise persistence of phase-locking requires a separate analysis, which we develop later.

\subsection{Deterministic dynamics and spectral representation}

A deterministic M\"{o}bius system on the two-dimensional torus can be written as
\begin{equation}\label{eq:det-Mobius-system}
    \begin{cases}
        \dot{\phi}=f(\phi,\tau),\quad f(\phi+2\pi,\tau)=f(\phi,\tau+2\pi)=f(\phi,\tau),\\ \dot{\tau}=\omega,
    \end{cases}
\end{equation}
where $\omega>0$ and the Fourier series of $f$ with respect to $\phi$ contains only the zeroth and first harmonics. We assume that the coefficients are real and smooth and fix the initial forcing phase at $\tau(0)=0$. The first recurrence map, i.e. the time-$T$ Poincar\'{e} map with $T=2\pi/\omega$, is an orientation-preserving M\"{o}bius transformation of the unit circle~\cite{marvel2009mobius}. All these transformations can be parametrized as
\begin{equation}\label{eq:Mobius-transformation}
    \mathcal M_{\zeta,w}(z)=\zeta\frac{z-w}{1-\overline{w}z},\quad |\zeta|=1,\quad |w|<1,\quad z=e^{i\phi}.
\end{equation}
They form the three-dimensional Lie group $\mathrm{PSU}(1,1)$. Its Lie algebra is spanned by the holomorphic vector fields
\begin{equation}\label{eq:Mobius-generators}
    \mathcal{Y}_1(z)=iz,\quad\mathcal{Y}_2(z)=i(z^2+1),\quad\mathcal{Y}_3(z)=z^2-1,
\end{equation}
whose brackets are closed as
\begin{equation}\label{eq:Mobius-commutators}
    [\mathcal{Y}_1,\mathcal{Y}_2]=-\mathcal{Y}_3,\qquad[\mathcal{Y}_1,\mathcal{Y}_3]=\mathcal{Y}_2,\qquad[\mathcal{Y}_2,\mathcal{Y}_3]=4\mathcal{Y}_1.
\end{equation}
Each function $\mathcal{Y}_r$ denotes the coefficient of $\partial_z$. Since $dz=iz\,d\phi$, these generators correspond to the phase fields $1$, $2\cos\phi$, and $2\sin\phi$. Absorbing the numerical factors, we introduce the normalized basis
\begin{equation}\label{eq:normalized-phase-basis}
    V_1(\phi)=1,\quad V_2(\phi)=\cos\phi,\quad V_3(\phi)=\sin\phi.
\end{equation}
The change of variable $\Phi=\tan(\phi/2)$ gives an affine coordinate on the projective circle $\mathbb{RP}^1$. The point $\phi=\pi$ corresponds to $\Phi=\infty$ and this is a chart boundary, not a singularity of the flow. The correspondence between the three representations is
\begin{center}
    \begin{tabular}{ccc}
    \toprule
    normalized disk field & phase field & projective field \\
    \midrule
    $iz$ & $1$ & $(1+\Phi^2)/2$ \\
    $i(z^2+1)/2$ & $\cos\phi$ & $(1-\Phi^2)/2$ \\
    $(z^2-1)/2$ & $\sin\phi$ & $\Phi$ \\
    \bottomrule
    \end{tabular}
\end{center}
Both $z$ and $\Phi$ satisfy the Riccati equations and are related by the Cayley transformation $z=(1+i\Phi)/(1-i\Phi)$. In the displayed order, these basis fields should not be identified with elliptic, parabolic, and hyperbolic types. For example, the parabolic phase field $1+\cos\phi$ corresponds to $i(z+1)^2/2$ and to the constant projective field $1$. Taking into account these facts, we can represent the general drift as follows
\begin{equation}\label{eq:gen-Mob-flow}
    f(\phi,\tau)=q_0(\tau)\left(1+\cos\phi\right)+q_1(\tau)\sin\phi+q_2(\tau)\left(1-\cos\phi\right),
\end{equation}
where $q_0$, $q_1$, and $q_2$ are smooth, real, $2\pi$-periodic functions. In the time variable $\tau=\omega t$, the corresponding Riccati equation for $\Phi$ is
\begin{equation}\label{eq:Riccati}
    \omega\frac{d\Phi}{d\tau}=q_0(\tau)+q_1(\tau)\Phi+q_2(\tau)\Phi^2.
\end{equation}
The examples of M\"{o}bius systems that we use in this study are listed in tab.~\ref{tab:Mobius-models}. The $\theta$-neuron is a standard model of type-I neuronal excitability~\cite{ermentrout1986bursting}, the RSJ model describes an overdamped Josephson junction~\cite{mccumber1968rsj,stewart1968rcsj} (here we have made $\pi/2$-shift of $\phi$ for convenience). The G2 model is a constructed extension of the forced $\theta$-neuron equation with input $I(\tau)=(1+\cos\tau)(B+A\cos2\tau)$. As we will see further, for generic parameters the normalization of its complexified critical curve, defined by $f(\tau,\phi)=0$, has genus two and it provides a higher-genus test case for the geometric and numerical analysis. Similar genus two models can arise from the Josephson diode~\cite{scheer2025tunable,borgongino2025biharmonic}.

\begin{table}[h!]
    \centering
    \small
    \setlength{\tabcolsep}{4pt}
    \begin{tabular}{ccc}
        \toprule
        model & $f(\phi,\tau)$ & origin \\
        \midrule
        $\theta$-neuron & $1-\cos\phi+(1+\cos\phi)(B+A\cos\tau)$ & neuronal models \\
        RSJ model & $-\cos\phi+B+A\cos\tau$ & superconductivity \\
        G2 model & $1-\cos\phi+(1+\cos\phi)(1+\cos\tau)(B+A\cos2\tau)$ & constructed example \\
        \bottomrule
    \end{tabular}
    \caption{Three illustrative M\"{o}bius systems.}
    \label{tab:Mobius-models}
\end{table}

Now, let us recall how phase-locking appears in this representation. A hyperbolic Poincar\'{e} map has one attracting and one repelling fixed point on $S^1$. An elliptic map is conjugate to a nontrivial rotation and has no fixed point on $S^1$, while a nontrivial parabolic map has one neutral fixed point. If the functions $q_0$, $q_1$ and $q_2$ depend on parameters, the open domains where the Poincar\'{e} map is hyperbolic are the robust phase-locked domains, usually called \emph{Arnold tongues}. The Poincar\'{e} rotation number,
\begin{equation}\label{eq:det-rho}
    \rho=\lim_{t\to\infty}\frac{\widetilde{\phi}(t)-\widetilde{\phi}(0)}{\omega t}=\lim_{n\to\infty}\frac{\widetilde{\phi}(nT)-\widetilde{\phi}(0)}{2\pi n},
\end{equation}
is an integer and is constant within each connected tongue, which was proven rigorously for the RSJ model in~\cite{buchstaber2010rotation}. In eq.~\eqref{eq:det-rho}, $\widetilde{\phi}\in\mathbb{R}$ is the continuous lift of the phase, so $\rho$ counts turns per forcing period. Elliptic maps describe the regions between tongues. Parabolic maps correspond to generic tongue boundaries. Also, notice that the identity map is an additional possibility at degenerate boundary points.

To connect this picture with spectral theory, suppose first that $q_2(\tau)\neq 0$ throughout the period. We set $K=q_1+\omega q_2'/q_2$ and $-\omega\psi'/\psi=q_2\Phi+K/2$, where a prime denotes $d/d\tau$. Then, eq.~\eqref{eq:Riccati} becomes
\begin{equation}\label{eq:Schrodinger}
    -\omega^2\frac{d^2\psi}{d\tau^2}+U(\tau,\omega)\psi=0,\quad U(\tau,\omega)=\frac{K(\tau,\omega)^2}{4}-q_0(\tau)q_2(\tau)-\frac{\omega}{2}\frac{dK(\tau,\omega)}{d\tau}.
\end{equation}
The potential $U$ is smooth, real, and periodic. If $q_2$ has zeros, this scalar reduction is only local and its apparent singularities need not be singularities of the M\"{o}bius dynamics. The matrix formulation introduced below remains regular and provides the global description.

\begin{figure}
    \centering
    \includegraphics[width=0.7\linewidth]{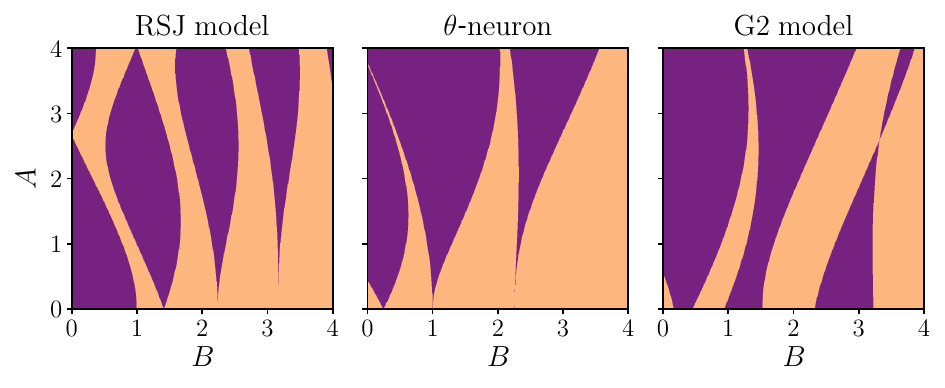}
    \caption{Bandgap structures of all three models: RSJ model, $\theta$-neuron, and G2 model. Arnold tongues (spectral gaps) are dark, gaps between tongues (spectral bands) are light.}
    \label{fig:bandgaps}
\end{figure}

For a globally regular scalar reduction, we consider the periodic Schr\"{o}dinger operator $H_\omega=-\omega^2d^2/d\tau^2+U(\tau,\omega)$ on $L^2(\mathbb{R})$. Its spectrum has a band-gap structure~\cite[Sec.~5.6]{teschl2012ode}, and eq.~\eqref{eq:Schrodinger} selects the energy $E=0$. Let $M$ be the monodromy matrix of this equation over one period. Then, $|\tr M|>2$ places zero in a spectral gap and gives a hyperbolic Poincar\'{e} map, whereas $|\tr M|<2$ places it in a band interior and gives an elliptic map. The case $|\tr M|=2$ corresponds to a band edge, including a closed gap when $M=\pm I$. The scalar and Riccati descriptions have conjugate projective monodromies because the change of variables is periodic. This is the known correspondence between Arnold tongues and spectral gaps. The introduced above Poincar\'{e} rotation number $\rho$ labels the gaps, according to the gap labeling theorem~\cite{bellissard1992gap,damanik2023gap}. As mentioned, throughout the discussion, we illustrate our findings using the three models, tab.~\ref{tab:Mobius-models}. The bandgap structures with $\omega=1$ for all three mentioned models are illustrated at fig.~\ref{fig:bandgaps}. For $0<\omega\ll1$, the WKB analysis becomes natural. The exact WKB method extends this analysis under the corresponding analytic assumptions~\cite{iwaki2025lectures}. 

\subsection{Noise preserves the M\"{o}bius structure}

We consider the stochastic generalization of~\eqref{eq:det-Mobius-system},
\begin{equation}\label{eq:stoch-Mobius-system}
    d\phi_t = f(\phi_t,\tau_t)\,dt+\sum_{r=1}^{3}\sqrt{2D_r}V_r(\phi_t)\circ dB_t^r,\quad d\tau_t=\omega\,dt.
\end{equation}
Here $B^1,B^2,B^3$ are independent standard Brownian motions, $D_r\geq0$ are constant noise intensities, we use notation $\bm{D}=(D_1,D_2,D_3)$, and $f(\phi,\tau)$ is given by eq.~\eqref{eq:gen-Mob-flow}. We use the three fixed generators~\eqref{eq:normalized-phase-basis} as independent noise channels. The first channel is additive, while the other two are multiplicative. The independent channels model fluctuations of the constant, cosine, and sine coefficients of the drift. Physical noise sources may instead act along fixed linear combinations of these fields, which also preserve the M\"{o}bius structure. For example, input fluctuations in the $\theta$-neuron act along $V_1+V_2=1+\cos\phi$, with the same Brownian motion multiplying both components. Since drift and diffusion fields belong to the same M\"{o}bius Lie algebra, this is a special case of a stochastic Lie-Scheffers system~\cite{lazarocami2009superposition}. We use the Stratonovich prescription, so the usual chain rule transforms eq.~\eqref{eq:stoch-Mobius-system} into
\begin{equation}\label{eq:stoch-Riccati}
    d\Phi_t=\left(q_0+q_1\Phi_t+q_2\Phi_t^2\right)dt
    +\sum_{r=1}^{3}\sqrt{2D_r}\,R_r(\Phi_t)\circ dB_t^r,
\end{equation}
where $q_j=q_j(\tau_t)$ and the normalized Riccati fields are
\begin{equation}\label{eq:normalized-Riccati-fields}
    R_1(\Phi)=\frac{1+\Phi^2}{2},
    \quad R_2(\Phi)=\frac{1-\Phi^2}{2},
    \quad R_3(\Phi)=\Phi.
\end{equation}
Equation~\eqref{eq:stoch-Riccati} lives on $\mathbb{RP}^1$ and an apparent pole in the affine coordinate $\Phi$ is only a projective chart crossing. We introduce traceless matrices,
\begin{equation}\label{eq:lift-matrices}
    A(\tau)=
    \begin{pmatrix} q_1/2 & q_0 \\ -q_2 & -q_1/2
    \end{pmatrix},\, K_1=\frac12
    \begin{pmatrix}0 & 1 \\ -1 & 0\end{pmatrix},\,
    K_2=\frac{1}{2}
    \begin{pmatrix}0 & 1\\ 1 & 0\end{pmatrix},\,
    K_3=\frac{1}{2}
    \begin{pmatrix}1 & 0\\0 & -1\end{pmatrix},
\end{equation}
which allows us to write down the equation for the fundamental matrix $G_t$,
\begin{equation}
    dG_t=A(\tau_t)G_t\,dt
    +\sum_{r=1}^{3}\sqrt{2D_r}\,K_rG_t\circ dB_t^r,
    \qquad G_0=I.
    \label{eq:matrix-Stratonovich-SDE}
\end{equation}
The following proposition is important to us:

\begin{proposition}[Pathwise M\"{o}bius closure]
For every fixed noise realization outside a null set and every $t\geq0$, eq.~\eqref{eq:stoch-Mobius-system} maps all initial phases, at the fixed initial forcing phase and driven by this same realization, by one orientation-preserving M\"{o}bius transformation. Equivalently, there are random parameters $\zeta_t$ and $w_t$, with $|\zeta_t|=1$ and $|w_t|<1$, such that
\begin{equation}\label{eq:pathwise-random-Mobius-map}
    z_t=\mathcal{M}_{\zeta_t,w_t}(z_0)
    =\zeta_t\frac{z_0-w_t}{1-\overline{w_t}z_0},
\end{equation}
\end{proposition}
where $z_t=e^{i\phi_t}$, $z_0=e^{i\phi_0}$ and $\phi_0$ is the initial phase.

\begin{proof}
All coefficients in eq.~\eqref{eq:matrix-Stratonovich-SDE} are traceless, then the ordinary determinant chain rule gives
\begin{equation}
    d\log\det G_t=\operatorname{tr}(G_t^{-1}\circ dG_t)=0,
\end{equation}
so $G_t\in\operatorname{SL}(2,\mathbb{R})$ pathwise. Its projective action is
\begin{equation}\label{eq:projective-action}
    \Phi_t=\frac{G_{11}(t)\Phi_0+G_{12}(t)}
    {G_{21}(t)\Phi_0+G_{22}(t)},
\end{equation}
which is precisely eq.~\eqref{eq:stoch-Riccati} with $\Phi_0=\tan(\phi_0/2)$. By the Stratonovich quotient rule and eq.~\eqref{eq:matrix-Stratonovich-SDE}, the process defined by eq.~\eqref{eq:projective-action} satisfies eq.~\eqref{eq:stoch-Riccati}. The Cayley conjugation turns $G_t$ into $\operatorname{SU}(1,1)$ matrix,
\begin{equation}
    \begin{pmatrix}
        \alpha_t&\beta_t\\
        \overline{\beta_t}&\overline{\alpha_t}
    \end{pmatrix},
    \qquad |\alpha_t|^2-|\beta_t|^2=1.
\end{equation}
With the standard action
\begin{equation}
    z\longmapsto
    \frac{\alpha_t z+\beta_t}
    {\overline{\beta_t}z+\overline{\alpha_t}},
\end{equation}
this matrix has the form in eq.~\eqref{eq:pathwise-random-Mobius-map}, where
\begin{equation}
    \zeta_t=\frac{\alpha_t}{\overline{\alpha_t}},
    \qquad
    w_t=-\frac{\beta_t}{\alpha_t}.
\end{equation}
The $\operatorname{SU}(1,1)$ identity implies $|w_t|<1$.
\end{proof}

\begin{figure}[h!]
    \centering
    \includegraphics[width=0.7\linewidth]{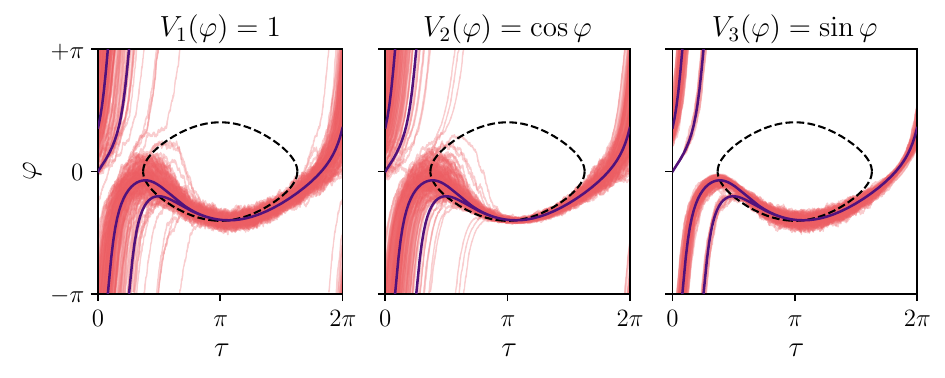}
    \caption{Influence of three noise channels $V_1$, $V_2$, and $V_3$ on the deterministic dynamics of overdamped Josephson junction (RSJ model) with $f(\phi,\tau)=-\cos\phi+B+A\cos\tau$, $A=0.5$, $B=0.8$, $\omega=0.2$. Dark solid curve corresponds to deterministic trajectory, thin red curves represent stochastic trajectories. The noise intensity is $D=0.01$ for each active channel. The dashed black curve represents frozen equilibrium $f(\phi,\tau)=0$.}
    \label{fig:trajectories}
\end{figure}

\begin{remark}[Why the Stratonovich form matters]
The It\^{o} equation equivalent to eq.~\eqref{eq:stoch-Mobius-system} has drift
\begin{equation}\label{eq:Ito-drift}
    f_{\text{Ito}}(\phi,\tau)=f(\phi,\tau)+\sum_{r=1}^{3}D_rV_r(\phi)V_r'(\phi).
\end{equation}
With our basis, eq.~\eqref{eq:normalized-phase-basis}, the correction is
\begin{equation}
    (D_3-D_2)\sin\phi\cos\phi,
\end{equation}
which is generally outside the three-dimensional M\"{o}bius algebra. This does not mean that the correctly converted It\^{o} equation has lost the M\"{o}bius flow: it describes the same process. The geometric closure is then no longer visible term by term. An It\^{o} equation written naively with the same drift $f$, but without the correction is a different model and need not preserve the M\"{o}bius structure.
\end{remark}

\noindent Fig.~\ref{fig:trajectories} illustrates how the three noise channels perturb deterministic trajectories. It is clear that equal channel intensities do not necessarily give equal local phase diffusion, since fields $V_r$ depend on the phase differently.

\subsection{Rotation number and probability current}

Pathwise M\"{o}bius closure does not imply that the deterministic bandgap structure survives unchanged. To study how noise affects it, we need a robust quantity, which is the rotation number. We lift the phase continuously to $\widetilde{\phi}_t\in\mathbb{R}$ by solving eq.~\eqref{eq:stoch-Mobius-system} with the $2\pi$-periodic coefficients extended to the real line. The lift is part of the dynamical data: an endpoint in $\mathrm{PSL}(2,\mathbb{R})$ alone does not retain the accumulated integer winding. For random circle maps, the rotation number can depend on the choice of lifts~\cite{rodrigues2015rotation}, but here the continuous stochastic flow fixes this choice. For example, $\exp(2\pi K_1)=-I$ acts trivially on the projective line, although the lifted phase has advanced by $2\pi$. Let $\widetilde{P}_n:\mathbb{R}\to\mathbb{R}$ be the lifted stochastic flow over the interval $[nT,(n+1)T]$. It is increasing and obeys the relation
\begin{equation}\label{eq:degree-one-property}
    \widetilde{P}_n(x+2\pi)=\widetilde{P}_n(x)+2\pi.
\end{equation}
The periodic deterministic coefficients and independent Brownian increments make $\widetilde{P}_0,\widetilde{P}_1,\ldots$ i.i.d. random maps. Their displacements along a trajectory are not independent, because the endpoint of one period is the initial phase of the next. The existence of the rotation number follows from the theory of stationary random circle homeomorphisms developed by Ruffino~\cite{ruffino2000sampling}. For the sake of completeness, we give a short proof adapted to the continuously lifted stochastic flow,

\begin{proposition}[Random Poincar\'{e} rotation number]
\label{prop:random-Poincare-rotation-number}
Assume that
\begin{equation}\label{eq:rotation-integrability}
    \mathbb{E}\left[\sup_{x\in[0,2\pi]}|\widetilde{P}_0(x)-x|\right]<\infty.
\end{equation}
Then, there is a deterministic number $\rho_{\bm{D}}\in\mathbb{R}$ such that, for almost every Brownian realization and every initial lift $\widetilde{\phi}_0$,
\begin{equation}\label{eq:almost-sure-rotation-number}
    \rho_{\bm{D}}=\lim_{n\to\infty}
    \frac{\widetilde{\phi}_{nT}-\widetilde{\phi}_0}{2\pi n}.
\end{equation}
\end{proposition}

\begin{proof}
Let $F_n=\widetilde{P}_{n-1}\circ\cdots\circ\widetilde{P}_0$ and denote the maximal and minimal displacements by
\begin{equation*}
    a_n=\sup_{x\in[0,2\pi]}[F_n(x)-x],\qquad b_n=\inf_{x\in[0,2\pi]}[F_n(x)-x].
\end{equation*}
Monotonicity and the degree-one property give us $0\leq a_n-b_n\leq 2\pi$. Next, the maximal displacement is subadditive under composition, while the minimal displacement is superadditive. The subadditive ergodic theorem and the ergodicity of the Brownian period shift give deterministic limits for $a_n/n$ and $b_n/n$. Their bounded difference makes these limits equal, proving eq.~\eqref{eq:almost-sure-rotation-number} simultaneously for all initial phases.
\end{proof}

\noindent The integrability condition holds for the present coefficients. Indeed, the monotonicity gives $\sup_x|\widetilde{P}_0(x)-x|\leq|\widetilde{P}_0(0)|+2\pi$, and the bounded It\^{o} coefficients imply $\mathbb{E}|\widetilde{P}_0(0)|<\infty$. The displacement during a remaining partial period is sublinear in time almost surely, so the same rotation limit holds at arbitrary times. We do not need an ellipticity assumption for this existence statement.

The rotation number~\eqref{eq:almost-sure-rotation-number} is directly related to the periodic probability current. To obtain a unique periodic statistical state and a simple spectral theory, we impose a uniform nondegeneracy in the phase variable. The local phase-diffusion coefficient is given by
\begin{equation}\label{eq:uniform-ellipticity}
    a_{\bm{D}}(\phi)=\sum_{r=1}^{3}D_rV_r(\phi)^2=D_1+D_2\cos^2\phi+D_3\sin^2\phi\geq d_*>0,\quad\phi\in S^1.
\end{equation}
Since $\min_{\phi}a_{\bm{D}}(\phi)=D_1+\min(D_2,D_3)$, this condition holds precisely when $D_1>0$ or both $D_2,D_3>0$. Notice that a single $V_2$ or $V_3$ channel does not satisfy eq.~\eqref{eq:uniform-ellipticity}. Such channels still preserve the M\"{o}bius structure and have a rotation number, but the spectral and large-deviation conclusions below require separate verification in these degenerate cases. The condition~\eqref{eq:uniform-ellipticity} ensures a smooth, strictly positive one-period phase kernel. The extended generator remains degenerate in $\tau$, whose evolution is deterministic. The backward generator on the extended torus is
\begin{equation}\label{eq:backward-generator}
    \mathcal{L}=\omega\partial_{\tau}+f(\phi,\tau)\partial_{\phi}+\sum_{r=1}^{3}D_rV_r(\phi)\partial_{\phi}\left[V_r(\phi)\partial_{\phi}\right].
\end{equation}
Under condition~\eqref{eq:uniform-ellipticity}, our process has a unique smooth positive periodic statistical state with density $p_*(\phi,\tau)$, which is normalized and solves the continuity equation,
\begin{equation}\label{eq:periodic-density}
    \int_0^{2\pi}d\phi\,p_*(\phi,\tau)=1,\quad \omega\partial_{\tau} p_*(\phi,\tau)
    =-\partial_{\phi} j_*(\phi,\tau),
\end{equation}
where the stationary periodic probability flux is
\begin{equation}\label{eq:probability-current}
    j_*(\phi,\tau)=f(\phi,\tau)p_*(\phi,\tau)-\sum_{r=1}^{3}D_rV_r(\phi)\partial_{\phi}\left[V_r(\phi)p_*(\phi,\tau)\right].
\end{equation}
Averaging the Fokker-Planck equation from~\eqref{eq:periodic-density} over one period shows that the mean flux defined by
\begin{equation}\label{eq:period-averaged-flux}
    \overline{j}=\frac{1}{2\pi}\int_0^{2\pi}d\tau\,j_*(\phi,\tau)
\end{equation}
is independent of $\phi$. Then, the asymptotic angular velocity is
\begin{equation}\label{eq:rotation-current-relation}
    \Omega_{\bm{D}}=\lim_{t\to\infty}\frac{\widetilde{\phi}_t-\widetilde{\phi}_0}{t}=\frac{1}{2\pi}\int_0^{2\pi}d\tau\int_0^{2\pi}d\phi\,j_*(\phi,\tau)=2\pi\overline{j}=\omega\rho_{\bm{D}},
\end{equation}
almost surely. Equivalently, we have $\rho_{\bm{D}}=T\overline{j}=\Omega_{\bm{D}}/\omega$. This is the precise relation between the lifted Poincar\'{e} rotation number and the stationary probability flux, consistent with the current formulation for periodically driven diffusions~\cite[Appendix B]{barato2018fluctuations}.

\subsection{Finite-time fluctuations and large deviations}

The limiting rotation number $\rho_{\bm{D}}$ is deterministic, while its finite-time estimators fluctuate. But its existence alone does not imply that the integer plateaus persist. In this subsection, we keep the noise intensities fixed and assume that condition eq.~\eqref{eq:uniform-ellipticity} holds. We introduce the lifted angular displacement, the empirical angular velocity, and the finite-time rotation number, respectively,
\begin{equation}\label{eq:finite-time-observables}
    \Theta_t=\widetilde{\phi}_t-\widetilde{\phi}_0,\quad\Omega_t=\frac{\Theta_t}{t},\quad\rho_n=\frac{\Theta_{nT}}{2\pi n}.
\end{equation}
By definition, the scaled cumulant generating function (SCGF) of $\Theta_t$ is
\begin{equation}\label{eq:SCGF}
    \lambda(k)=\lim_{t\to\infty}
    \frac{1}{t}\log\mathbb{E}\,e^{k\Theta_t}.
\end{equation}
The expectation value can be taken from any fixed initial phase or from the periodic stationary law. Under condition~\eqref{eq:uniform-ellipticity}, the limit is the same. A precise one-period construction can be obtained from the tilted kernel,
\begin{equation}\label{eq:tilted-one-period-kernel}
    (\mathcal{P}_k h)(\phi)=\mathbb{E}_{\phi,0}\left[e^{k(\widetilde{\phi}_T-\widetilde{\phi}_0)}h(\phi_T\bmod 2\pi)\right].
\end{equation}
The uniform ellipticity makes $\mathcal{P}_k$ a strongly positive compact operator on $C(S^1)$, and its dominant eigenvalue is simple. Therefore, we can write
\begin{equation}\label{eq:SCGF-spectral-radius}
    \lambda(k)=\frac{1}{T}\log\operatorname{spr}(\mathcal{P}_k),
\end{equation}
where $\operatorname{spr}(\mathcal{P}_k)$ is the spectral radius. In this setting, $\lambda(k)$ is differentiable (indeed analytic) for real $k$ and $\lambda'(0)=\Omega_{\bm{D}}=\omega\rho_{\bm{D}}$. We can obtain the same construction using a tilted periodic differential operator. For autonomous circle diffusion, the corresponding current fluctuations were studied in~\cite{tsobgninyawo2016current}, while the periodic forcing leads to a Floquet problem~\cite{barato2018fluctuations}. We introduce the tilted derivative $\mathcal{D}_k=\partial_{\phi}+k$, so the ordered Stratonovich form of the tilted generator is
\begin{equation}\label{eq:tilted-generator}
    \mathcal{L}_k=\omega\partial_{\tau}+f(\phi,\tau)\mathcal{D}_k+\sum_{r=1}^{3}D_rV_r(\phi)\mathcal{D}_kV_r(\phi)\mathcal{D}_k,
\end{equation}
where multiplication by $V_r$ and application of $\mathcal{D}_k$ are composed as is. Formally, we can write $\mathcal{L}_k=e^{-k\widetilde{\phi}}\mathcal{L}e^{k\widetilde{\phi}}$ on periodic functions of the projected phase. The number $\lambda(k)$ is the principal Floquet exponent: a positive function $h_k$, periodic in both variables, satisfies $\mathcal{L}_k h_k=\lambda(k)h_k$. Restriction of this function to $\tau=0$ gives the positive eigenfunction of $\mathcal{P}_k$ with the eigenvalue $e^{T\lambda(k)}$.

The bounded coefficients give exponential tightness, while the simple dominant eigenvalue gives a finite differentiable SCGF on the real axis. The G\"{a}rtner-Ellis theorem~\cite[Theorem~2.3.6]{dembo2010deviations} gives us the current large-deviation principle (LDP),
\begin{equation}\label{eq:current-LDP}
    \mathbb{P}(\Omega_t\simeq u)\asymp e^{-tI_\Omega(u)}, \quad I_\Omega(u)=\sup_{k\in\mathbb{R}}\,\{ku-\lambda(k)\},
\end{equation}
where the probability notation is shorthand for the usual large-deviation upper and lower bounds, and $I_\Omega$ a good rate function. Since $\rho_n=\Omega_{nT}/\omega$, the finite-time Poincar\'{e} rotation estimator obeys the equivalent per-period form,
\begin{equation}\label{eq:rotation-LDP}
    \mathbb{P}(\rho_n\simeq\varrho)\asymp e^{-nI_\rho(\varrho)},\quad I_\rho(\varrho)=TI_\Omega(\omega\varrho).
\end{equation}
The obtained equations~\eqref{eq:almost-sure-rotation-number}, \eqref{eq:rotation-current-relation}, and \eqref{eq:rotation-LDP} separate three logically distinct statements: the pathwise existence of the infinite-time rotation number, its identification with the typical angular velocity, and the exponentially small probability of atypical finite-time windings. We emphasize that this long-time LDP is taken at fixed noise intensities $\bm{D}$. A related homological-current approach to large deviations for elliptic diffusions on compact manifolds is developed in~\cite{galkin2024homology}. It does not estimate how close $\rho_{\bm{D}}$ is to a deterministic integer as the noise vanishes. That question concerns weak-noise phase-slips and their Freidlin-Wentzell costs, which we consider below.

\subsection{Melting of Arnold tongues}

Beyond the finite-time LDP, we distinguish two weak-noise questions: how accurately the infinite-time rotation number of Proposition~\ref{prop:random-Poincare-rotation-number} follows a deterministic tongue and how long a trajectory remains locked. The first question concerns the net phase-slip current, while the second concerns the total slip activity. We use these quantities to formulate persistence and observable melting.

Let us start from the action derivation. Inside a deterministic tongue with rotation number $\ell\in\mathbb{Z}$, we fix $\omega>0$ and relative channel strengths $\nu_r\geq0$, then set
\begin{equation}\label{eq:weak-noise-scale}
    D_r=D\nu_r,\quad \varepsilon=\frac{D}{\omega},\quad D\to0.
\end{equation}
For fixed scalar coefficients $V_r(\phi)$ of the phase generators, we define
\begin{equation}\label{eq:fw-diffusion-coefficient}
    a(\phi)=\sum_r\nu_rV_r(\phi)^2,\quad b(\phi,\tau)=\frac{f(\phi,\tau)}{\omega}.
\end{equation}
In slow time $\tau=\omega t$, the equation for phase and its Freidlin-Wentzell (FW) action for an absolutely continuous lifted path $\gamma$ are given by
\begin{gather}
    d\phi=b(\phi,\tau)\,d\tau+\sqrt{2\varepsilon}\sum_r \sqrt{\nu_r}V_r(\phi)\circ dW^r_\tau, \label{eq:fw-slow-time-sde} \\
    \mathcal{I}_{[\tau_0,\tau_1]}[\gamma]=\inf_{\bm{u}\in L^2}\left\{\frac{1}{4}\int_{\tau_0}^{\tau_1}d\tau\,\sum_r u_r(\tau)^2:\ \gamma'=b(\gamma,\tau)+\sum_r \sqrt{\nu_r}V_r(\gamma)u_r\right\}. \label{eq:fw-control-action}
\end{gather}
We assume that inaccessible paths have infinite cost. For $a>0$, minimization over the controls gives
\begin{equation}\label{eq:fw-path-action}
    \boxed{\mathcal{I}_{[\tau_0,\tau_1]}[\gamma]=\frac{1}{4}\int_{\tau_0}^{\tau_1}d\tau\,\frac{[\gamma'(\tau)-f(\gamma(\tau),\tau)/\omega]^2}{a(\gamma(\tau))}.}
\end{equation}
This cost weights deviations from the deterministic motion by the available noise. The probabilities of small path neighborhoods scale as $\exp(-\mathcal{I}/\varepsilon)$~\cite{freidlin2012random,venttsel1970perturbations}. The factor $1/4$ reflects the convention $\sqrt{2\varepsilon}$, and the $O(\varepsilon)$ Stratonovich-to-It\^{o} drift correction leaves the leading action unchanged.

Let $\phi_s$ be the attracting locked lift and $\phi_u^-<\phi_s<\phi_u^+=\phi_u^-+2\pi$ its adjacent repelling separators, all satisfying $\phi(\tau+2\pi)=\phi(\tau)+2\pi\ell$. For the deterministic parameters set $\bm{p}$, the forward and backward barriers are
\begin{equation}\label{eq:fw-slip-barriers}
    \boxed{S_{\pm}^{\text{FW}}(\bm{p})=\inf_{\substack{\tau_0\in[0,2\pi),\ \tau_1>\tau_0\\ \gamma(\tau_0)=\phi_s(\tau_0),\ \gamma(\tau_1)=\phi_u^\pm(\tau_1)}}\mathcal{I}_{[\tau_0,\tau_1]}[\gamma].}
\end{equation}
The admissible paths stay between the separators until exit. The infimum optimizes the forcing phase and unrestricted duration, which possibly requires an infinite-duration limit. At fixed deterministic parameters, we choose sufficiently small disjoint neighborhoods of the attracting lifts $\phi_s(\tau)+2\pi k$, $k\in\mathbb{Z}$, which are contained in their respective deterministic basins and independent of $D$. We count a \emph{completed slip}, when the trajectory first reaches a neighborhood closest to its last recorded lift. The lift index is then updated, so repeated crossings of a separator are not counted separately. Under nondegenerate noise, this completion has the same leading action as reaching the separator, but its prefactor can differ. Each completed slip changes the lift index by $\pm1$ and contributes a signed turn to the accumulated winding.

The dependence of $S_{\pm}^{\text{FW}}$ on $\omega$ and the fixed channels is implicit. For the physical-time cost $\int dt\,(\dot\gamma-f)^2/(4a)$, the corresponding barriers satisfy simple relations
\begin{equation}\label{eq:fw-time-normalization}
    R_{\pm}^{\text{FW}}=\omega S_{\pm}^{\text{FW}},\quad \frac{S_{\pm}^{\text{FW}}}{D/\omega}=\frac{R_{\pm}^{\text{FW}}}{D}.
\end{equation}
It is important that $D/\omega$ alone does not determine escape, when $\omega$ varies.

\noindent\textbf{Rotation number accuracy and locking time.} We assume smooth periodic coefficients, hyperbolic locked orbits, and $a\geq a_0>0$ on the basin closure, including its separators. Degenerate channels require separate accessibility checks using eq.~\eqref{eq:fw-control-action}. Let $N_{\pm}(n)$ count completed positive and negative slips during $n$ forcing periods. Under the corresponding ergodic assumptions, we define their asymptotic rates per forcing period by the almost-sure limits,
\begin{equation}\label{eq:completed-slip-count-rates}
    \Gamma_{\pm}=\lim_{n\to\infty}\frac{N_{\pm}(n)}{n},\quad \Gamma_{\Sigma}=\Gamma_{+}+\Gamma_{-}.
\end{equation}
The weak-noise escape asymptotics give us the following logarithmic law
\begin{equation}\label{eq:phase-slip-rates}
    \lim_{\varepsilon\to0}\varepsilon\log\Gamma_{\pm}=-S_{\pm}^{\text{FW}},\quad \Gamma_{\pm}=C_{\pm}(\bm{p},\varepsilon)\exp\left[-\frac{S_{\pm}^{\text{FW}}}{\varepsilon}\right],\quad \varepsilon\log C_{\pm}\to0.
\end{equation}
This law determines the exponential scale, but does not by itself provide an Eyring-Kramers prefactor. In particular, the prefactors for periodic-orbit escape need not be noise independent~\cite{berglund2014passage}. The accumulated winding differs from $\ell n+N_{+}(n)-N_{-}(n)$ by a bounded remainder. Dividing by $n$ and taking the long-time limit therefore gives us
\begin{equation}\label{eq:noisy-rotation-slip-correction}
    \rho_{\bm{D}}-\ell=\Gamma_{+}-\Gamma_{-},\quad |\rho_{\bm{D}}-\ell|\leq\Gamma_{\Sigma},
\end{equation}
where $\rho_{\bm{D}}$ counts net turns per forcing period. This identity does not require a Poisson approximation for the completed-slip process. In a compact interior of the tongue $\mathcal{K}$ at fixed $\omega$ and fixed relative channel strengths, uniform hyperbolicity and uniform escape estimates yield
\begin{equation}\label{eq:exponential-tongue-persistence}
    S_{\mathcal{K}}=\inf_{\bm{p}\in\mathcal{K}}\min\bigl(S_{+}^{\text{FW}}(\bm{p}),S_{-}^{\text{FW}}(\bm{p})\bigr)>0,\quad\limsup_{\varepsilon\to0}\varepsilon\log\left[\sup_{\bm{p}\in\mathcal{K}}|\rho_{\bm{D}}(\bm{p})-\ell|\right]\leq-S_{\mathcal{K}}.
\end{equation}
The deviation is beyond all algebraic orders and generically nonzero, although symmetry may cancel the net current despite ongoing slips. With a uniform subexponential bound $C_{\mathcal{K}}(\varepsilon)$ on $C_{+}+C_{-}$, rotation number accuracy $\eta>0$ is ensured by the following expression,
\begin{equation}\label{eq:infinite-time-small-noise-condition}
    C_{\mathcal{K}}(\varepsilon)e^{-S_{\mathcal{K}}/\varepsilon}\leq\eta,\quad \boxed{\frac{D}{\omega}\leq\frac{S_{\mathcal{K}}}{\log[C_{\mathcal{K}}(\varepsilon)/\eta]}.}
\end{equation}
Here, the logarithm must be positive, and the condition is implicit when $C_{\mathcal{K}}$ depends on the noise. Without prefactor control, eq.~\eqref{eq:exponential-tongue-persistence} guarantees us $\sup_{\mathcal{K}}|\rho_{\bm{D}}-\ell|\leq e^{-(S_{\mathcal{K}}-h)/\varepsilon}$ for each $0<h<S_{\mathcal{K}}$ and sufficiently small $\varepsilon$, but does not give a numerical threshold.

We say that a relaxed trajectory is \emph{metastably locked} over $N$ forcing periods when its completed-slip probability is small. When $\Gamma_{\Sigma}\ll 1$ and relaxation is fast compared with slip waiting times, the rare-slip approximation gives the mean waiting time and survival law,
\begin{gather}
    N_{\mathrm{lock}}\simeq\Gamma_{\Sigma}^{-1},\quad T_{\mathrm{lock}}\simeq\frac{2\pi}{\omega}N_{\mathrm{lock}}, \label{eq:metastable-locking-time} \\
    \mathbb{P}\{\text{at least one completed slip during $N$ periods}\}\simeq1-e^{-N\Gamma_{\Sigma}}. \label{eq:phase-slip-probability}
\end{gather}
Weak noise produces rare phase slips whose rates are exponentially small inside deterministic locking regions. We call phase-locking \emph{metastable}, when the probability of a completed slip remains small over a prescribed observation time. Its lifetime can therefore be exponentially long, while the asymptotic rotation number remains exponentially close to its deterministic locked value. This exponential survival law is valid after relaxation on the many-period waiting-time scale. Moreover, the slip probability within a period can still depend on the forcing phase. For the tolerated probability $0<\alpha<1$, we set $L_{\alpha}=-\log(1-\alpha)$. Within the rare-slip approximation, the phase-locking criterion is
\begin{equation}\label{eq:metastable-locking-condition}
    N\Gamma_{\Sigma}\leq L_{\alpha}.
\end{equation}
Writing $\Gamma_{\Sigma}=C_*(\varepsilon)e^{-S_*^{\text{FW}}/\varepsilon}$ with $S_*^{\text{FW}}=\min(S_{+}^{\text{FW}},S_{-}^{\text{FW}})$ and a subexponential $C_*$ yields
\begin{equation}\label{eq:metastable-small-noise-condition}
    \boxed{\frac{D}{\omega}\lesssim\frac{S_*^{\text{FW}}(\bm{p})}{\log[C_*(\varepsilon)N/L_{\alpha}]}},\quad C_*(\varepsilon)N>L_{\alpha}.
\end{equation}
We see that rotation number accuracy constrains the net current, while metastable phase-locking constrains the total activity. Then a sufficient activity tolerance is $q_{\text{tol}}=\eta$ for the accuracy and $q_{\text{tol}}=L_{\alpha}/N$ for the phase-locking, or their minimum for both. We associate observable melting with the loss of the chosen accuracy or survival criterion.

\begin{figure}[h!]
    \centering
    \includegraphics[width=0.9\linewidth]{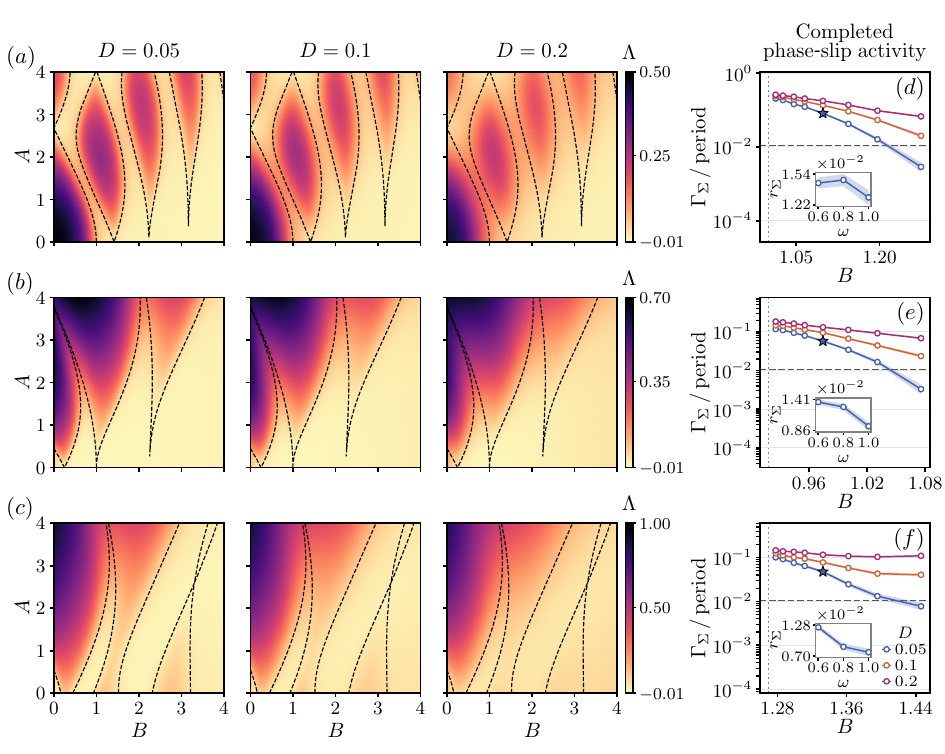}
    \caption{(a)-(c) Melting tongues for different additive noise intensities for RSJ model, $\theta$-neuron, and G2-model. The color represents top Lyapunov exponent. The black dashed curves show deterministic tongue boundaries; (d)-(f) Total completed-phase slip activity $\Gamma_{\Sigma}$ per forcing period along $A=1$ cuts through tongues with rotation numbers $\ell=1,\,2,\,2$ respectively. Circles mark measurements, shaded bands demonstrate $95\%$ confidence intervals. Horizontal dashed line shows tolerance $\Gamma_{\text{tol}}=-\log(1-\alpha)/N_{\text{obs}}\simeq1.05\times10^{-2}$ , with $N_{\text{obs}}=10$ and $\alpha=0.1$. Insets show the total activity per physical time, $r_{\Sigma}=\omega\Gamma_{\Sigma}/(2\pi)$, at $D=0.05$, following the same tongue at the fixed fractional position $u_0=0.15$ between its deterministic boundaries. Stars mark the corresponding points at $\omega=1$.}
    \label{fig:melted-tongues}
\end{figure}

We can visualize noisy tongues and their melting using the top Lyapunov exponent $\Lambda$ of the associated random matrix product. Let $M_n\in SL(2,\mathbb{R})$ be the independent, identically distributed propagators over one forcing period $T=2\pi/\omega$, and let $\Pi_N=M_N\cdots M_1$. We normalize the exponent per physical time and write down
\begin{equation}\label{eq:tongue-matrix-product-exponent}
    \widehat\Lambda_N=\frac{1}{NT}\log\|\Pi_N\|\xrightarrow[N\to\infty]{\mathrm{a.s.}}\Lambda,
\end{equation}
where the matrix norm is the Euclidean operator norm. Let $m_{\bm{D}}$ be the law of $M_n$, and let $\mu_{\bm{D}}$ be a forward stationary distribution of projective directions selecting the top matrix exponent. With finite logarithmic moments, the Furstenberg formula~\cite{furstenberg1963products} allows us to express $\Lambda$ as
\begin{equation}\label{eq:tongue-furstenberg-matrix}
    \Lambda=\frac{1}{T}\int dm_{\bm{D}}(M)\int d\mu_{\bm{D}}([v])\,\log\frac{\|Mv\|}{\|v\|},
\end{equation}
where $[v]$ is the line spanned by $v\neq0$. To relate this quantity to phase dynamics, we define $\Lambda_{\phi}=\lim_{t\to\infty}t^{-1}\log|\partial\phi_t/\partial\phi_0|$ for trajectories driven by the same noise realization. Then, in the projective convention of eq.~\eqref{eq:projective-action}, the unit vector $v_{\phi}=(\sin(\phi/2),\cos(\phi/2))^T$ represents the phase $\phi$. The determinant-one property gives us
\begin{equation}\label{eq:tongue-matrix-phase-exponents}
    \left|\frac{\partial\phi_{NT}}{\partial\phi_0}\right|=\|\Pi_Nv_{\phi_0}\|^{-2},\quad \boxed{\Lambda_{\phi}=-2\Lambda.}
\end{equation}
The limiting relation holds for a forward direction selecting the top exponent. The finite-time identity involves growth along that direction, whereas eq.~\eqref{eq:tongue-matrix-product-exponent} uses the operator norm. If matrix growth is reported per forcing period instead, its exponent is $\Lambda_{\text{per}}=T\Lambda$ and $\Lambda_{\phi}=-2\Lambda_{\text{per}}/T$. Thus, positive matrix growth corresponds to contraction of nearby phase trajectories driven by the same noise. The same exponent also admits a representation through the phase process. Under ergodicity, the continuous-time Furstenberg-Khasminskii formula~\cite{khasminskii2012stability} can be written as
\begin{equation}\label{eq:tongue-furstenberg-phase}
    \begin{gathered}
    \boxed{\Lambda_{\phi}=\frac{1}{2\pi}\int_0^{2\pi}d\tau\int_0^{2\pi}d\phi\,p_{*}(\phi,\tau)Q_{\bm{D}}(\phi,\tau),}\\
    Q_{\bm{D}}(\phi,\tau)=\partial_{\phi} f(\phi,\tau)+\sum_rD_rV_r(\phi)V_r''(\phi),
    \end{gathered}
\end{equation}
where primes denote phase derivatives and $p_*(\phi,\tau)$ is the normalized periodic stationary density of~\eqref{eq:periodic-density}. In eq.~\eqref{eq:tongue-furstenberg-phase}, the second term comes from converting the Stratonovich equation for the logarithm of the tangent flow to It\^{o} form. For additive noise it vanishes, while for nonconstant channels averaging $\partial_{\phi} f$ alone generally gives the wrong exponent. After discarding an initial transient of duration $t_b$, an observation time $\mathcal{T}$ provides the equivalent phase estimator,
\begin{equation}\label{eq:tongue-furstenberg-estimator}
    \widehat\Lambda_{\phi}(\mathcal{T})=\frac{1}{\mathcal{T}}\int_{t_b}^{t_b+\mathcal{T}}dt\,Q_{\bm{D}}(\phi_t,\omega t)\xrightarrow[\mathcal{T}\to\infty]{\mathrm{a.s.}}\Lambda_{\phi}.
\end{equation}
The It\^{o} martingale in the logarithmic tangent equation, divided by $\mathcal{T}$, vanishes in this limit. Thus, $-\widehat\Lambda_{\phi}(\mathcal{T})/2$ provides a check on the matrix estimate of $\Lambda$, although the finite-time estimates do not necessarily coincide. At zero noise, $\Lambda>0$ in hyperbolic tongue interiors and $\Lambda=0$ in the elliptic regions of the M\"obius return map, as should be. With noise, a map of $\Lambda$ visualizes the deformation and blurring of the Arnold tongues. Positive matrix growth can also occur between deterministic tongues and coexist with completed slips, so its sign alone does not define a melting boundary. The rotation and survival criteria above give the corresponding quantitative test of metastable phase-locking. Examples of Arnold tongues melting are shown at fig.~\ref{fig:melted-tongues}(a)-(c). In all plots $\Lambda$ represents the top Lyapunov exponent, obtained by the Furstenberg formula. The phase-slip activity for all three models is shown at fig.~\ref{fig:melted-tongues}(d)-(f).

\noindent\textbf{Rounding near a regular boundary.} What happens near a regular tongue boundary? At fixed $\omega$, let $\delta>0$ measure transverse distance into a saddle-node boundary. In a signed local lifted-phase coordinate $x$, the return map $P(x,\delta)=x+c_1\delta-c_2x^2+\cdots$, with $c_1,c_2>0$, admits the local diffusion approximation
\begin{equation}\label{eq:noisy-saddle-node-normal-form}
    dx=(c_1\delta-c_2x^2)\,dn+\sqrt{2D_{\mathrm{eff}}}\,dW_n,\quad D_{\mathrm{eff}}=\varkappa\frac{D}{\omega},
\end{equation}
where $n$ counts periods and the coefficients are evaluated at the boundary. To make the projected noise explicit, let $\phi_c(\tau)$ be the coalescing locked orbit. Linear propagation of the noise over one period gives variance $2\varepsilon\varkappa$ to leading order, where
\begin{equation}\label{eq:tongue-projected-noise}
    \varkappa=\int_0^{2\pi}d\tau\,a(\phi_c(\tau))G(2\pi,\tau)^2,\quad G(\tau_2,\tau_1)=\exp\left[\int_{\tau_1}^{\tau_2}ds\,\partial_{\phi} b(\phi_c(s),s)\right].
\end{equation}
Here $x$ is the lifted-phase displacement, up to an orientation change. A further coordinate rescaling changes $c_1$, $c_2$, and $\varkappa>0$ consistently. The stable and unstable states are $x_s=\sqrt{c_1\delta/c_2}$ and $x_u=-\sqrt{c_1\delta/c_2}$. The potential $U_{\text{sn}}=c_2x^3/3-c_1\delta x$ tell us that the barrier height is
\begin{equation}\label{eq:saddle-node-barrier}
    \Delta U_{\text{sn}}=\frac43\frac{(c_1\delta)^{3/2}}{\sqrt{c_2}},\quad S_*^{\text{FW}}=\frac{\Delta U_{\text{sn}}}{\varkappa}[1+o(1)],\quad \delta\downarrow0.
\end{equation}
Assuming that $\Delta U_{\text{sn}}\gg D_{\mathrm{eff}}$, Kramer's formula gives the completed-slip rate~\cite{risken1996fokkerplanck},
\begin{equation}\label{eq:saddle-node-slip-rate}
    \Gamma_{\mathrm{slip}}\simeq\frac{\sqrt{c_1c_2\delta}}{\pi}\exp\left[-\frac{4(c_1\delta)^{3/2}}{3\sqrt{c_2}\,D_{\mathrm{eff}}}\right].
\end{equation}
We assume that passage through this bottleneck is followed by deterministic relaxation to the neighboring lift and that the opposite slip barrier remains positive. Then $\Gamma_{\Sigma}\simeq\Gamma_{\mathrm{slip}}$ near the boundary in the activated regime. For the full system, higher-order action terms matter whenever their contribution divided by $\varepsilon$ is appreciable.

To obtain the intrinsic rounding scale without extrapolating Kramers' formula, let us set $x=(D_{\mathrm{eff}}/c_2)^{1/3}y$ and $s=c_2^{2/3}D_{\mathrm{eff}}^{1/3}n$. Then, we can write down
\begin{equation}\label{eq:intrinsic-tongue-rounding}
    dy=(\lambda-y^2)\,ds+\sqrt2\,dW_s,\quad \lambda=\frac{c_1\delta}{c_2^{1/3}D_{\mathrm{eff}}^{2/3}},\quad\boxed{\delta_{\mathrm{round}}=\frac{c_2^{1/3}}{c_1}D_{\mathrm{eff}}^{2/3}.}
\end{equation}
The crossover passage rate is of order $c_2^{2/3}D_{\mathrm{eff}}^{1/3}$ and need not admit a rare-slip Poisson description~\cite{hathcock2021rates}. If $q_{\text{tol}}$ selects a point in the Kramers region, the \emph{observable} melting boundary, defined by the condition $\Gamma_{\mathrm{slip}}(\delta_{\mathrm{melt}})=q_{\text{tol}}$, satisfies
\begin{equation}\label{eq:tongue-melting-layer}
    \boxed{\delta_{\mathrm{melt}}\simeq\frac{c_2^{1/3}}{c_1}\left[\frac{3D_{\mathrm{eff}}}{4}\log\left(\frac{\sqrt{c_1c_2\delta_{\mathrm{melt}}}}{\pi q_{\text{tol}}}\right)\right]^{2/3}.}
\end{equation}
The logarithm must be large and positive, so the tolerance must lie well below the crossover rate. At fixed positive tolerance, that rate eventually falls below tolerance as noise decreases. Then, a positive inward melting width does not need to exist, although the intrinsic $2/3$ scale remains meaningful. The local small-noise criteria are
\begin{equation}\label{eq:local-small-noise-condition}
    \frac{\Delta U_{\text{sn}}}{D_{\mathrm{eff}}}\gg1,\quad \frac{\Delta U_{\text{sn}}}{D_{\mathrm{eff}}}\gtrsim\log\left(\frac{\sqrt{c_1c_2\delta}}{\pi q_{\text{tol}}}\right).
\end{equation}
When tolerance-defined layers exist at both edges of a tongue, they leave a resolved metastable core if their widths are small compared with the tongue width in the same parameter coordinate. Exponentially narrow tongues can require exponentially small noise when the normal-form coefficients remain controlled. Canard persistence additionally requires control of noise amplification along the repelling branch. Degenerate tips, vanishing projected noise, and joint slow-forcing limits require separate analysis. In particular, periodically driven escape can exhibit different intermediate scaling regimes~\cite{ryvkine2004scaling}.

\subsection{Adler equation as a simple example}

To illustrate the proposed above statements, we use the simplest possible example, which is the noisy Adler equation~\cite{adler1946locking},
\begin{equation}\label{eq:adler-noisy}
    d\phi=(A-B\sin\phi)\,dt+\sqrt{2D}\,dW_t,\quad B>0,\quad |A|<B.
\end{equation}
This autonomous system exhibits phase-locking at $\ell=0$ and makes the distinction between net current and slip activity explicit. An arbitrary observation interval $T=2\pi/\omega$ supplies the preceding normalization, with no physical forcing frequency implied. Let us denote $a_A=A/B$, $\chi=\arcsin a_A$, and $\lambda_A=\sqrt{B^2-A^2}$. The attracting lift and adjacent repellers are
\begin{equation}\label{eq:adler-locked-states}
    \phi_s=\chi,\quad \phi_u^+=\pi-\chi,\quad \phi_u^-=-\pi-\chi.
\end{equation}
For the tilted potential $U_A(\phi)=-A\phi-B\cos\phi$, completing the square gives
\begin{equation}\label{eq:adler-action-derivation}
    \frac{1}{4}\int_{t_i}^{t_f}dt\,(\dot\gamma+\partial_{\phi}U_A)^2=U_A(\gamma_f)-U_A(\gamma_i)+\frac{1}{4}\int_{t_i}^{t_f}dt\,(\dot\gamma-\partial_{\phi}U_A)^2,
\end{equation}
where $\partial_{\phi}U$ is evaluated along $\gamma$. The optimal uphill path follows the reversed drift $\dot\gamma=U_A'$ in the infinite-duration limit, and its cost is the potential difference. Therefore, the barriers are
\begin{equation}\label{eq:adler-exact-barriers}
    \boxed{\Delta U_{\pm}=2B\left[\sqrt{1-a_A^2}+a_A\arcsin a_A\mp\frac{\pi a_A}{2}\right],\quad S_{\pm}^{\text{FW}}=\frac{\Delta U_{\pm}}{\omega}.}
\end{equation}
Since $\Delta U_{-}-\Delta U_{+}=2\pi A$, the smaller barrier is
\begin{equation}\label{eq:adler-minimum-barrier}
    \Delta U_*=\min(\Delta U_{+},\Delta U_{-})=2B\left[\sqrt{1-a_A^2}-|a_A|\arccos|a_A|\right].
\end{equation}
Both endpoint curvatures have magnitude $\lambda_A$. For $D\ll\Delta U_*$, the physical-time rates $r_{\pm}$ and per-interval rates $\Gamma_{\pm}$ are
\begin{equation}\label{eq:adler-kramers-rates}
    r_{\pm}\simeq\frac{\lambda_A}{2\pi}e^{-\Delta U_{\pm}/D},\quad \boxed{\Gamma_{\pm}=T r_{\pm}\simeq\frac{\lambda_A}{\omega}e^{-\Delta U_{\pm}/D}.}
\end{equation}
\begin{figure}[h!]
    \centering
    \includegraphics[width=0.8\linewidth]{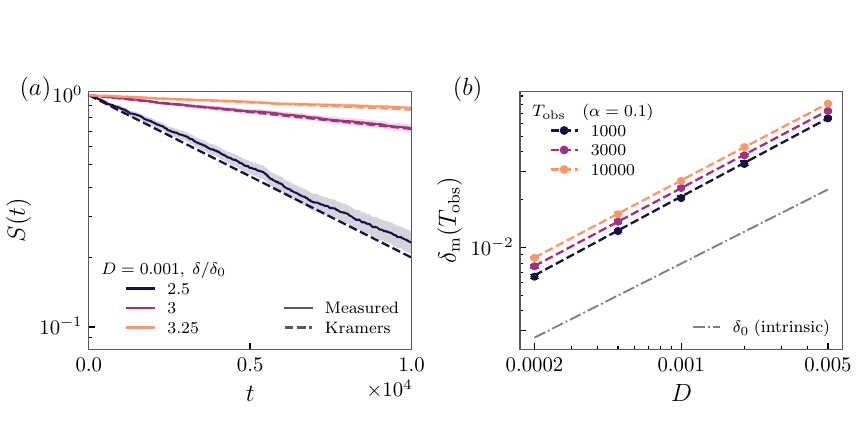}
    \caption{Adler equation simulations with $B=1$. (a) Empirical survival probability for forward threshold crossing $S(t)=\mathbb{P}(t_{\mathrm{slip}}>t)$ for Adler equation with initial condition $\phi_s=\arcsin(1-\delta)$ with $D=10^{-3}$ for several values $\delta/\delta_0$. Solid curves show simulation results, dashed curves show the asymptotic Kramers prediction. The first-slip time is defined by the first hit of the $3\pi/2$ phase. Shading indicated $95\%$ intervals; (b) Melting widths $\delta_{\text{m}}$, defined by $\mathbb{P}(t_{\text{slip}}\leq T_{\text{obs}})=\alpha$ with $\alpha=0.1$ for several observation times. Circles denote interpolation probability crossing. Dashed curves satisfy $r_{+}T_{\text{obs}}=-\ln(1-\alpha)$ on the activated branch $\Delta U_{+}/D\geq 5$. The dash-dotted curve shows the intrinsic rounding scale $\delta_0=(D^2/2)^{1/3}$ without a fitted amplitude.}
    \label{fig:adler-melting}
\end{figure}
When $T(r_{+}+r_{-})\ll1$, these rates describe rare completed slips per observation interval. The mean rotation number and locking time become
\begin{equation}\label{eq:adler-rotation-locking}
    \rho_D\simeq\frac{\lambda_A}{\omega}\left(e^{-\Delta U_{+}/D}-e^{-\Delta U_{-}/D}\right),\qquad
    N_{\mathrm{lock}}\simeq\frac{\omega}{\lambda_A\left(e^{-\Delta U_{+}/D}+e^{-\Delta U_{-}/D}\right)}.
\end{equation}
For either activity tolerance above, a sufficient leading-order criterion is
\begin{equation}\label{eq:adler-small-noise}
    \frac{2\lambda_A}{\omega}e^{-\Delta U_*/D}\leq q_{\text{tol}},\quad \boxed{D\lesssim\frac{\Delta U_*}{\log[2\lambda_A/(\omega q_{\text{tol}})]},}
\end{equation}
requiring a positive logarithm and $D\ll\Delta U_*$. Beyond the weak-noise approximation, we can solve the stationary Fokker-Planck equation and obtain explicit expression for the current~\cite{risken1996fokkerplanck}:
\begin{equation}\label{eq:adler-exact-current}
    \rho_D=T j_D,\quad j_D=\frac{D\left(1-e^{-2\pi A/D}\right)}{\displaystyle\int_0^{2\pi}dx\,e^{-U_A(x)/D}\int_x^{x+2\pi}dy\,e^{U_A(y)/D}}.
\end{equation}
This gives an exact benchmark for all $D>0$. Since $j_D$ counts net rotations per physical time, the mean angular velocity is $2\pi j_D$. The weak-noise limit recovers eq.~\eqref{eq:adler-rotation-locking}. At $A=0$, symmetry gives $\rho_D=0$ for every $D>0$, yet
\begin{equation}\label{eq:adler-symmetric-locking}
    \Gamma_{\Sigma}\simeq\frac{2B}{\omega}e^{-2B/D},\quad N_{\mathrm{lock}}\simeq\frac{\omega}{2B}e^{2B/D}.
\end{equation}
Therefore, an integer mean rotation coexists with a finite waiting time for loss of trajectory locking. The Furstenberg formula also tell us that $\Lambda_{\phi}=-B\langle\cos\phi\rangle$. At $A=0$, the stationary density is proportional to $e^{B\cos\phi/D}$, and we can write down the top Lyapunov exponent explicitly,
\begin{equation}\label{eq:adler-furstenberg-exponent}
\begin{gathered}
    \Lambda_{\phi}=-B\frac{I_1(B/D)}{I_0(B/D)}<0,\quad \Lambda=\frac{B}{2}\frac{I_1(B/D)}{I_0(B/D)}>0,\\
    \Lambda_{\phi}=-B+\frac{D}{2}+O(D^2/B),\quad \Lambda=\frac{B}{2}-\frac{D}{4}+O(D^2/B),\quad D/B\to 0,
\end{gathered}
\end{equation}
where $I_n$ denotes the modified Bessel function of the first kind of order $n$. Thus, contraction persists despite completed slips, and its leading change is algebraic in noise while the slip rate is exponentially small. Near the threshold, $\delta=B-|A|\downarrow0$ at fixed $B$ gives
\begin{equation}\label{eq:adler-boundary-scaling}
    \Delta U_*=\frac{4\sqrt2}{3\sqrt B}\delta^{3/2}\left[1+O(\delta/B)\right],\quad \lambda_A\sim\sqrt{2B\delta},\quad\delta_{\mathrm{round}}=(B/2)^{1/3}D^{2/3}.
\end{equation}
The last expression is the characteristic scale obtained by rescaling the local normal form. It is independent of the arbitrary observation frequency, which enters only the criteria per observation interval. Outside phase-locking, $\rho_0=\operatorname{sgn}(A)\sqrt{A^2-B^2}/\omega$ for $|A|>B$. In fig.~\ref{fig:adler-melting}(a), the empirical survival probability to observe phase-slip after time $t$ is shown for different ratios $\delta/\delta_0$, $\delta_0=(D^2/2)^{1/3}$. The melting widths $\delta_{\text{m}}$ are shown in fig.~\ref{fig:adler-melting}(b). For this figure, $t_{\mathrm{slip}}=\inf\{t>0:\phi(t)=3\pi/2\}$ denotes the first forward threshold-crossing time of the unwrapped phase. This threshold lies beyond the forward repeller and before the neighboring attractor. The stopping rule used in fig.~\ref{fig:adler-melting} measures forward threshold crossing, whereas the completed-slip definition above requires arrival near the neighboring attracting lift. In the activated regime near $A=B$, these events have the same leading activation barrier, and the subsequent relaxation time is negligible compared to the escape waiting time. However, their finite-time survival probabilities can differ outside this regime. Backward slips are exponentially suppressed near this positive locking threshold.

\section{WKB method builds deterministic skeleton}
\label{sec:WKB-deterministic-skeleton}

In the slow-fast limit, $0<\omega\ll1$, we can apply the WKB method and construct the deterministic Arnold tongues from the $\mathrm{SL}(2,\mathbb{R})$ lift of eq.~\eqref{eq:Riccati},
\begin{equation}\label{eq:WKB-deterministic-lift}
    \omega G_{\det}'=\omega\frac{dG_{\det}}{d\tau}=A(\tau)G_{\det}, \quad G_{\det}(0)=I.
\end{equation}
Here $\tau=\omega t$ is the slow time, $I$ is the $2\times2$ identity matrix, and the subscript ``$\det$'' means ``deterministic''. The real $2\pi$-periodic matrix $A(\tau)$ is given by eq.~\eqref{eq:lift-matrices}. Its fundamental matrix $G_{\det}$ defines the one-period monodromy,
\begin{equation}\label{eq:WKB-deterministic-monodromy}
    M_{\det}=G_{\det}(2\pi)\in\mathrm{SL}(2,\mathbb{R}).
\end{equation}
The WKB method relates its elliptic, hyperbolic and parabolic types to transport on a spectral curve. The classical WKB actions give the leading exponential scales and determine the leading-order bandgap structure, while the exact WKB includes the corrections needed to resolve exponentially narrow spectral features~\cite{delabaere1997semiclassical,iwaki2014clusters}.

\subsection{From Riccati dynamics to Floquet bands}

Let $q_0,q_1,q_2$ be the real $2\pi$-periodic coefficients of $\omega\Phi'=q_0+q_1\Phi+q_2\Phi^2$, with the projective phase coordinate $\Phi=\tan(\phi/2)$. These coefficients are independent of $\omega$. In a chart with $q_2\neq 0$, the scalarization~\eqref{eq:Schrodinger},
\begin{equation}\label{eq:WKB-scalarization}
    K=q_1+\omega\frac{q_2'}{q_2}, \quad -\omega\frac{\psi'}{\psi}=q_2\Phi+\frac{K}{2},
\end{equation}
gives the Schr\"{o}dinger equation for the auxiliary scalar function $\psi$,
\begin{equation}\label{eq:WKB-Schrodinger-equation}
    -\omega^2\psi''+U(\tau,\omega)\psi=0,
\end{equation}
with the periodic potential of form
\begin{equation}\label{eq:WKB-potential-expansion}
\begin{split}
    U&=U_0+\omega U_1+\omega^2U_2, \quad U_0=\frac{\Delta_0}{4}, \quad \Delta_0=q_1^2-4q_0q_2,\\
    U_1&=\frac{1}{2}\left(q_1\frac{q_2'}{q_2}-q_1'\right), \quad U_2=\frac14\left(\frac{q_2'}{q_2}\right)^2-\frac{1}{2}\left(\frac{q_2'}{q_2}\right)'.
\end{split}
\end{equation}
We see that the leading term $U_0$ is determined by the frozen Riccati discriminant $\Delta_0$ and the terms $U_1$ and $U_2$ are derivative corrections from scalarization. Zeros of $q_2$ require another scalar chart, although the matrix equation remains regular when its coefficients are regular.

Let $M_{\psi}$ be the one-period transfer matrix for $(\psi,\omega\psi')^{T}$, and let $L(\tau)$ be the invertible gauge taking a vector solution of the lift to these scalar coordinates. Then, the monodromy of the Shcr\"{o}dinger equation is
\begin{equation}\label{eq:WKB-gauge-monodromy}
    M_{\psi}=L(2\pi)M_{\det}L(0)^{-1}.
\end{equation}
A periodic gauge makes the two matrices conjugate, as should be. Otherwise, we should restore the endpoint factors and any scalar-chart transitions. In a periodic gauge, the Floquet half-discriminant $\mathcal{F}$ is
\begin{equation}\label{eq:WKB-Floquet-discriminant}
    \mathcal{F}=\frac{1}{2}\tr M_{\det}=\frac{1}{2}\tr M_\psi.
\end{equation}
For real coefficients, Floquet theory gives~\cite{teschl2012ode,alexandrov2026phaselock}
\begin{equation}\label{eq:WKB-Floquet-classification}
\begin{array}{ccl}
|\mathcal{F}|<1&\Longleftrightarrow&\text{elliptic monodromy, a band between tongues},\\[1mm]
|\mathcal{F}|>1&\Longleftrightarrow&\text{hyperbolic monodromy, a gap inside a tongue},\\[1mm]
\mathcal{F}=\pm1&\Longleftrightarrow&\text{parabolic or scalar monodromy}.
\end{array}
\end{equation}
The hyperbolic map has attracting and repelling fixed phases modulo one period, and a generic tongue boundary is parabolic. The exceptional case $M_{\det}=\pm I$ gives the identity projective map and corresponds to a closed gap in the usual Hill energy problem. When a control parameter enters as an additive energy, the classification is the usual Hill bandgap structure. Otherwise, it describes the zero-energy problem in control-parameter space. The half-discriminant determines the monodromy type, but it does not determine the integer winding of a locked orbit, i.e. rotation number. In a hyperbolic region, let $v_s$ be an eigenvector of $M_{\det}$ with eigenvalue $\mu_s$ satisfying $|\mu_s|>1$, its projective line is attracting. We propagate this line as $[G_{\det}(\tau)v_s]$ and continuously unwrap its phase $\phi_s(\tau)$ using $\Phi=v_1/v_2=\tan(\phi/2)$. Then, the rotation number that labels the tongue is
\begin{equation}\label{eq:WKB-tongue-label}
    \ell=\frac{\phi_s(2\pi)-\phi_s(0)}{2\pi}\in\mathbb{Z}.
\end{equation}
This means that the integer label tongue requires the lifted phase evolution over the period in addition to the endpoint monodromy. 

\subsection{Slow curve and classical WKB curve}

The slow curve of the Riccati system is directly related to the classical WKB curve, so-called classical spectral curve. Indeed, at $\omega=0$, the slow (or critical) curve for Riccati system is defined by $\mathcal{C}_0:\,q_2\Phi^2+q_1\Phi+q_0=0$ with branches
\begin{equation}\label{eq:WKB-critical-branches}
    \Phi_{\pm}=\frac{-q_1\pm\sqrt{\Delta_0}}{2q_2}.
\end{equation}
Its normally hyperbolic real branches support nearby slow manifolds for small positive $\omega$~\cite{fenichel1979geometric,kuehn2015timescales,kristiansen2023review}. The leading WKB quadratic differential and its two-sheeted spectral cover~\cite{nikolaev2024geometry} are given by
\begin{equation}\label{eq:WKB-classical-quadratic-differential}
    \mathcal{Q}_0=U_0\,d\tau^2=\frac{\Delta_0}{4}\,d\tau^2,
\end{equation}
\begin{equation}\label{eq:WKB-spectral-curve}
    \Sigma_0:\quad y^2=U_0=\frac{\Delta_0}{4}, \quad \lambda_0=y\,d\tau, \quad \lambda_0^2=\mathcal{Q}_0.
\end{equation}
Here, $y$ denotes the leading WKB momentum and $\lambda_0$ its action differential. The substitution $y=q_2\Phi+q_1/2$, $\Phi=(y-q_1/2)/q_2$ maps $\mathcal{C}_0$ to $\Sigma_0$ and gives $2y=q_1+2q_2\Phi=\partial_\Phi(q_0+q_1\Phi+q_2\Phi^2)$. It is easy to see that the frozen lift has eigenvalues $\pm y$, so on a real critical branch $2y<0$ gives attraction and $2y>0$ gives repulsion, with transverse rate $2y/\omega$ in slow time. Apart from the common zeros of $q_0$, $q_1$, $q_2$, changes in the projective chart identify normalizations of the complexified slow curve and the spectral curve. Resolving the algebraic singularities, we obtain the smooth curves. The real locus over real $\tau$ gives us the critical curve. Simple zeros of $\Delta_0$ are WKB turning points and generic folds of the real critical curve, when they lie on the real forcing cycle~\cite{kristiansen2024turning}.

\subsection{Finite Fourier drives give finite genus}

\begin{figure}[h!]
    \centering
    \includegraphics[width=0.8\linewidth]{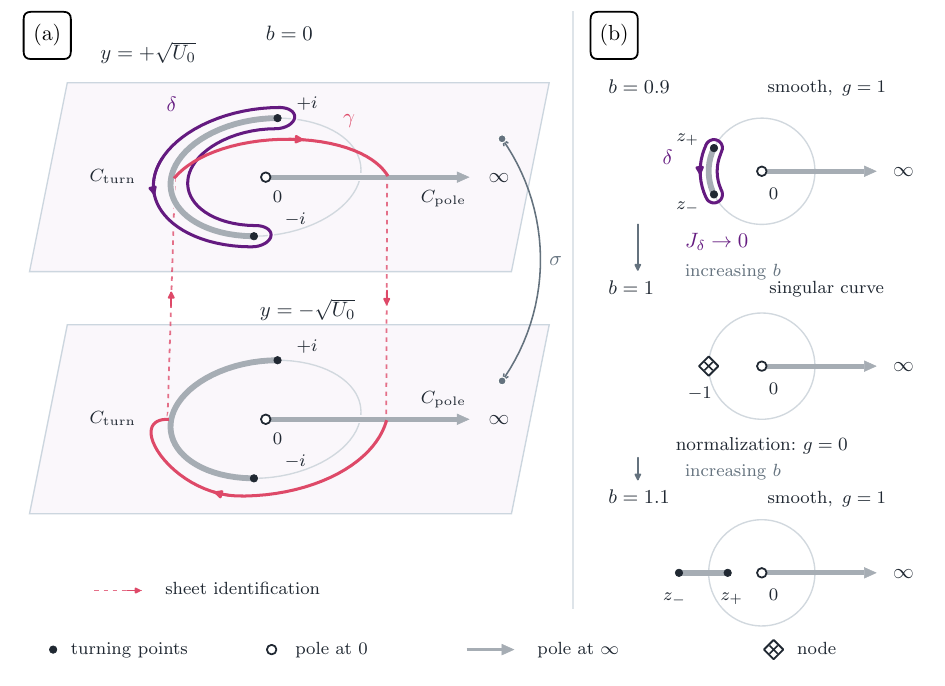}
    \caption{Classical WKB double cover for the Mathieu equation, with $U_0=-(B+A\cos\tau)$ and $b=B/A$. (a) With $z=e^{i\tau}$, the cover is branched at the turning points $z_\pm=-b\pm\sqrt{b^2-1}$ and at the fixed poles $0,\infty$. Crosswise gluing along the chosen cuts gives a genus-one surface. The sheet exchange is $\sigma:(z,y)\mapsto(z,-y)$, and $\gamma\circ\delta=1$. (b) As $b\to1$, the turning points meet at $z=-1$ and the barrier cycle $\delta$ contracts, with $J_{\delta}\to0$. The limiting curve is nodal and its normalization has genus zero. For $b>1$, the turning points separate off the unit circle and the smooth curve again has genus one. The cuts indicate the gluing convention.}
    \label{fig:wkb-double-cover}
\end{figure}

The two-sheeted cover introduced above~\eqref{eq:WKB-spectral-curve} is typically represented by an elliptic or hyperelliptic curve with a certain genus. The most general, but tractable case is the finite genus. The case of genus-1 is well-known, while the case of a larger genus requires more attention~\cite{fischbach2019wkb,ture2026quantumgeo}. It is pretty obvious that the finite genus requires more than smooth periodicity. Let us assume that $q_0$, $q_1$, $q_2$ are trigonometric polynomials of maximal harmonic order $n$, and let $M\geq 1$ be the actual Fourier degree of their non-constant discriminant after possible cancelations, so we can write
\begin{equation}\label{eq:WKB-discriminant-Fourier-degree}
    \Delta_0(\tau)=\sum_{k=-M}^{M}d_ke^{ik\tau}, \quad d_{-M}d_M\neq0, \quad M\leq2n.
\end{equation}
The Fourier coefficients satisfy $d_{-k}=\overline{d_k}$ for real parameters. With $z=e^{i\tau}$ (do not be confused with $z$ from discussion generators in~\eqref{eq:Mobius-generators}!), the complexified forcing cylinder becomes the punctured $z$-plane, so the Riccati discriminant and the WKB differential are
\begin{equation}\label{eq:WKB-z-form}
    \Delta_0(z)=z^{-M}P_{2M}(z), \quad \lambda_0^2=-\frac{P_{2M}(z)}{4z^{M+2}}\,dz^2,
\end{equation}
where $P_{2M}$ is a polynomial of degree $2M$ with nonzero constant term. We compactify the cover by adding the points $z=0$ and $z=\infty$, and use its normalization to define the genus $g$. If $P_{2M}$ has distinct roots and the cover is connected, the branch points are these roots, together with $0$ and $\infty$ when $M$ is odd. The Riemann-Hurwitz formula~\cite{forster1981riemann} gives
\begin{equation}\label{eq:WKB-genus}
    g(\Sigma_0)=
    \begin{cases}M-1,&M\ \text{even},\\M,&M\ \text{odd}.\end{cases}
\end{equation}
Thus, $g\geq 1$ in this nondegenerate case and generically we can have $M=2n$ and $g=2n-1$. The picture is a little bit clear with the change $u=\tan(\tau/2)$,
\begin{equation}\label{eq:WKB-hyperelliptic-form}
    \Delta_0=\frac{R_{4n}(u)}{(1+u^2)^{2n}}, \quad Y=2(1+u^2)^ny, \quad Y^2=R_{4n}(u),
\end{equation}
where $R_{4n}$ has degree at most $4n$. If it has degree $4n$ and no repeated roots, the curve has genus $2n-1$. It is elliptic for $g=1$ and hyperelliptic for $g\geq 2$. For our $\theta$-neuron model, we deal with Mathieu equation and the WKB double cover is shown in fig.~\ref{fig:wkb-double-cover}(a).

More generally, let $B_{\text{br}}$ count the odd-order zeros and poles of the meromorphic quadratic differential $\mathcal{Q}_0$ on a compact base of genus $g_{\text{base}}$. For a connected double cover, $g=2g_{\text{base}}-1+B_{\text{br}}/2$. On the sphere this reduces to $g=(B_{\text{br}}-2)/2$. Collisions and Fourier cancellations can lower the genus of the normalized curve. This classical finite-genus property does not imply that the Hill operator has only finitely many open spectral gaps, as was illustrated by the Mathieu equation example in~\cite{dunne2017mathieu}.

\subsection{Cycles, actions and quantum periods}

On the smooth compact spectral curve, we choose a symplectic basis of closed cycles~\cite{forster1981riemann},
\begin{equation}\label{eq:WKB-cycle-basis}
    \{\alpha_i,\beta_i\}_{i=1}^{g}, \quad \alpha_i\circ\beta_j=\delta_{ij}, \quad \alpha_i\circ\alpha_j=\beta_i\circ\beta_j=0,
\end{equation}
where symbol $\circ$ denotes the oriented intersection number and $\delta_{ij}$ is the Kronecker delta. For an oriented cycle $\gamma$, the classical action is defined by
\begin{equation}\label{eq:WKB-classical-period}
    J_{\gamma}=\oint_{\gamma}\lambda_0.
\end{equation}
Its real part controls exponential transport, while its imaginary part controls oscillatory phase through the quantity $J_{\gamma}/\omega$. Our cycles avoid poles of the differential. The connection problem may also require small loops around poles and relative paths with specified endpoints and normalizations~\cite{iwaki2014clusters}. Such physical connection cycles need not form a symplectic basis.

The WKB ansatz $\psi=\exp\{\omega^{-1}\int d\tau\,S\}$ gives $S=\omega\psi'/\psi$, which satisfies $S^2+\omega S'=U$. Expanding two formal solutions of this equation,
\begin{equation}\label{eq:WKB-momentum-series}
    S^{(\pm)}\sim\sum_{k\geq 0}\omega^kS_k^{(\pm)}, \quad S_0^{(\pm)}=\pm y,
\end{equation}
we obtain the recurrence relation
\begin{equation}\label{eq:WKB-momentum-recurrence}
    S_k^{(\pm)}=\frac{U_k-\left(S_{k-1}^{(\pm)}\right)'}{2S_0^{(\pm)}}-\frac{1}{2S_0^{(\pm)}}\sum_{j=1}^{k-1}S_j^{(\pm)}S_{k-j}^{(\pm)}, \quad k\geq1.
\end{equation}
The sum is empty for $k=1$ and for our case $U_k=0$ for $k\geq 3$. Away from turning points and poles, this determines the sheet-odd differential and its formal quantum periods~\cite{iwaki2014clusters,nikolaev2023riccati},
\begin{equation}\label{eq:WKB-quantum-period}
    \begin{split}
    S_{\text{odd}}&=\frac{S^{(+)}-S^{(-)}}{2}, \quad \lambda_{\text{WKB}}=S_{\text{odd}}\,d\tau\sim\lambda_0+\sum_{k\geq1}\omega^k\lambda_k,\\
    \Pi_{\gamma}^{\text{form}}&=\oint_{\gamma}\lambda_{\text{WKB}}\sim J_{\gamma}+\sum_{k\geq1}\omega^kJ_{\gamma}^{(k)},
    \end{split}
\end{equation}
where the contours avoid poles of the corresponding WKB differentials, including those arising at turning points, and
\begin{equation}\label{eq:WKB-lambda-coefficients}
    \lambda_k=\frac{S_k^{(+)}-S_k^{(-)}}{2}\,d\tau, \quad J_{\gamma}^{(k)}=\oint_{\gamma}\lambda_k.
\end{equation}
In particular, for our potential $U=U_0+\omega U_1+\omega^2U_2$, we have
\begin{equation}\label{eq:WKB-first-lambda-corrections}
    \lambda_1=\frac{U_1}{2y}\,d\tau, \quad \lambda_2=\left(\frac{U_2}{2y}-\frac{U_1^2}{8y^3}+\frac{y''}{4y^2}-\frac{3(y')^2}{8y^3}\right)d\tau.
\end{equation}
The sheet-even part supplies the amplitude,
\begin{equation}\label{eq:WKB-even-momentum}
    S_{\text{even}}=\frac{S^{(+)}+S^{(-)}}{2}=-\frac{\omega}{2}\frac{S_{\text{odd}}'}{S_{\text{odd}}}, \quad \psi_{\pm}\sim\frac{1}{\sqrt{S_{\text{odd}}}}\exp\left\{\pm\frac{1}{\omega}\int d\tau\,S_{\text{odd}}\right\}.
\end{equation}
We label the local sheets by the signs $+$ and $-$. The terms ``odd'' and ``even'' describe symmetry under the sheet exchange $\sigma:(\tau,y)\mapsto(\tau,-y)$, which interchanges $S^{(+)}$ and $S^{(-)}$. Therefore, $S_{\text{odd}}$ changes sign, while $S_{\text{even}}$ remains unchanged. When $U_1=0$, the potential is even in $\omega$ and the formal solutions satisfy $S^{(-)}(\tau,\omega)=-S^{(+)}(\tau,-\omega)$, so the sheet-odd part contains only even powers of $\omega$, while the sheet-even part contains only odd powers, providing $\lambda_{2m+1}=0$ for all $m\geq0$. The minus sign in eq.~\eqref{eq:WKB-scalarization} means that the WKB solution with leading momentum $+y$ approaches the critical branch with $q_2\Phi+q_1/2=-y$. By definition, a quantum period is an action corrected to all formal orders. Its exponential is the formal Voros symbol,
\begin{equation}\label{eq:WKB-formal-Voros-symbol}
    \mathcal{V}_{\gamma}^{\text{form}}=\exp\!\left[\Pi_{\gamma}^{\text{form}}/\omega\right].
\end{equation}
Our double-cover $\Sigma_0$ provides the classical geometry, whereas the quantum periods and their Voros symbols encode the transport around each cycle, including the classical action and its formal corrections. They provide the cycle data for the monodromy construction, together with amplitudes and connection factors~\cite[Secs.~1.1 and 1.4]{iwaki2025lectures}. However, near a shrinking cycle, we need to use the continuation and uniform approximations discussed next.

\subsection{Degenerations and continuation of periods}
\label{subsec:WKB-degenerations}

We assume analytic dependence on the complexified control parameters $\boldsymbol{\xi}$, with a meromorphic extension in the forcing coordinate. Then, in a regular scalar chart, finite turning-point collisions away from poles lie on the classical degeneration locus,
\begin{equation}\label{eq:WKB-degeneration-locus}
    \mathcal{D}_{\text{cl}}=\left\{\boldsymbol{\xi}:\ \exists\tau_*\text{ with }U_0(\tau_*,\boldsymbol{\xi})=\partial_\tau U_0(\tau_*,\boldsymbol{\xi})=0\right\}.
\end{equation}
For simple turning points and at fixed Fourier degree with $d_{-M}d_M\neq0$, this locus is simply $\operatorname{Disc}_zP_{2M}=0$, where $\operatorname{Disc}_z$ denotes the polynomial discriminant. We also should distinguish persistent multiple zeros from the parameter induced collisions of simple zeros. They are resolved when we compute the normalized curve, but still remain multiple turning points of the differential equation. In contrast, collisions with poles and degree changes should be examined on the compactified cover. For illustration of shrinking cycle, we use the Mathieu equation, see fig.~\ref{fig:wkb-double-cover}(b). 

Let us also discuss our models. For the $\theta$-neuron model, the finite turning points are simple, when $A\neq0$ and $B\neq\pm A$ (it is precisely Mathieu equation). For the RSJ model, $U_0=[1-(B+A\cos\tau)^2]/4$, and finite turning points are simple when $A\neq 0$ and $B\pm A$ equals $\pm 1$, which was used in~\cite{alexandrov2026duck}. We should treat zeros of a sclar-chart coefficients with the described above chart transitions. Our G2 model has an additional persistent multiple turning point. Indeed, the leading order potential is
\begin{equation}\label{eq:WKB-G2-double-turning-point}
    U_0=-(1+\cos\tau)(B+A\cos2\tau)=-\frac{B+A}{2}x^2+O(x^4),\quad x=\tau-\pi,
\end{equation}
so $\tau=\pi$ is a double turning point whenever $B+A\neq 0$. For $A\neq 0$ and $B\neq\pm A$, the remaining turning points are simple. Normalization of the classical curve does not remove this double turning point. Since $q_1=0$ and $q_2=1$, we have $U_1=U_2=0$. Its leading local comparison is the Weber equation~\eqref{eq:WKB-Weber-subleading} at $\nu=0$, with $a=-(B+A)/2$ and consistent choices of its roots. The global connection product requires the corresponding Weber connection factor, see~\cite{sueishi2021anomaly}.

Let us consider this generic double turning point. Let $\varepsilon$ be a local parameter transverse to $\mathcal{D}_{\text{cl}}$, with the collision at $\varepsilon=0$. With $x=\tau-\tau_*$, the local potential can be represented as
\begin{equation}\label{eq:WKB-double-turning-point}
    U_0=a x^2-b\varepsilon+O(x^3,\varepsilon x,\varepsilon^2), \quad a=\frac12\partial_\tau^2U_0\neq0, \quad b=-\partial_\varepsilon U_0\neq0,
\end{equation}
where the derivatives are evaluated at the collision. The cycle $\delta$ encircling the merging pair contracts and its classical action behaves as
\begin{equation}\label{eq:WKB-vanishing-period}
    J_{\delta}(\varepsilon)=c\varepsilon+O(\varepsilon^2)\longrightarrow0, \quad c\neq 0,
\end{equation}
where the constant $c$ depends on the local coefficients and the orientation of the cycle. For one non-separating pinch, the limiting genus is $g-1$. Every nearby smooth curve still has the genus $g$, including real parameter values on either side of the collision.

Periods are compared by continuing both the differential and its cycles along a specified detour in complex parameter space. Near this ordinary node, the Picard-Lefschetz formula~\cite{vassiliev2002picardlefschetz} gives us the following relation,
\begin{equation}\label{eq:WKB-period-continuation}
    J_{\gamma}(\varepsilon)=H_{\gamma}(\varepsilon)+\frac{n_{\gamma\delta}}{2\pi i}J_{\delta}(\varepsilon)\log\varepsilon,
\end{equation}
where $H_{\gamma}$ is analytic and $n_{\gamma\delta}$ is the signed integer determined by the cycle intersection and orientation convention. Equivalently, it is fixed by the counterclockwise parameter monodromy $\gamma\mapsto\gamma+n_{\gamma\delta}\delta$. Thus, the action remains finite, while the derivative $\partial_{\varepsilon} J_{\gamma}$ has a logarithmic singularity if $n_{\gamma\delta}\neq0$ and the logarithm is defined on the chosen branch. Since we have $\log(-|\varepsilon|\pm i0)=\log|\varepsilon|\pm i\pi$, the two detours differ by a multiple of the vanishing period. Therefore, continuation can mix the real and imaginary parts of an action. For several collisions, the corresponding transformations are composed in the order specified by the path. Simply speaking, this continuation changes the WKB transport through
\begin{equation}\label{eq:WKB-period-phase}
    e^{J_{\gamma}/\omega}=e^{\re J_{\gamma}/\omega}e^{i\im J_{\gamma}/\omega}.
\end{equation}
In general, a transported cycle and a cycle chosen to describe the new real turning-point configuration may have different reality and positivity properties. For example, continuation of a positive barrier action need not remain a positive suppression action~\cite{misumi2025degenerate,misumi2026nondegenerate}.

When $|J_{\delta}|$ is comparable to $\omega$ or smaller, separate Airy connections at the two turning points cease to be uniform. As we have mentioned, the Weber approximation~\cite{olver1975turning,bucciotti2023anharmonic,kristiansen2025eigenvalue}, expressed in parabolic-cylinder functions, treats the merging pair together. To illustrate this case, let us assume that $U_1$, $U_2$ and their derivatives remain bounded as $\varepsilon\to 0$ in a regular scalar chart. For $\varepsilon=O(\omega)$, we introduce rescaling,
\begin{equation}\label{eq:WKB-Weber-scaling}
    x=\omega^{1/2}a^{-1/4}s, \quad \Psi(s)=\psi(\tau_*+\omega^{1/2}a^{-1/4}s), \quad U_j^*=U_j(\tau_*,0),
\end{equation}
with consistent choices of the roots of $a$. Then, our local equation becomes
\begin{equation}\label{eq:WKB-Weber-subleading}
    -\frac{d^2\Psi}{ds^2}+\left[s^2-\nu+R(s,\varepsilon,\omega)\right]\Psi=0, \quad \nu=\frac{b\varepsilon-\omega U_1^*-\omega^2U_2^*}{\omega\sqrt{a}},
\end{equation}
with $R$ contains the remaining Taylor terms of full potential and $R=O(\omega^{1/2})$ for bounded $s$ and bounded $\varepsilon/\omega$. If we drop $R$, we will obtain the frozen quadratic comparison equation. This means that $\omega U_1^*$ changes the leading Weber parameter, while bounded $\omega^2U_2^*$ contributes at order $\omega$ in the scaled equation. The last term matters when the residual detuning $b\varepsilon-\omega U_1^*$ is resolved on the scale $O(\omega^2)$, or when corrections of order $\omega$ for the scaled equation are needed. Keeping only $U_2^*$ does not give a complete approximation at this accuracy. To finalize, the higher Taylor terms of $U_0$, the variation of $U_1$ can affect the connection problem and should be considered consistently.

It happens that the order counting changes near a scalar-chart singularity. For example, if $q_2=x^r h(x)$ with integer $r\geq1$ and analytic $h$ satisfying $h(0)\neq0$, then eq.~\eqref{eq:WKB-potential-expansion} gives us
\begin{equation}\label{eq:WKB-U2-chart-pole}
    U_2=\frac{r(r+2)}{4x^2}+O(x^{-1}).
\end{equation}
If $U_0\sim ax^2$ at the same point, the rescaling $x=\omega^{1/2}a^{-1/4}s$ turns $\omega^2U_2$ into the leading inverse-square term $r(r+2)/(4s^2)$ in the scaled equation. So, we cannot neglect it, and the regular Weber approximation is not applicable in that chart. When the lift is regular, we use a regular scalar chart where available, otherwise we use the matrix equation. For genuine poles, we need a local model that includes their singular terms. The inverse-square term above is already part of the exact scalarization, so no additional correction should be inserted into $U$ by hand.

These observations concern the uniform approximation of the full equation. The classical degeneration locus $\mathcal{D}_{\text{cl}}$ and the classical periods are still defined by $U_0$, but a double zero of $U_0+\omega U_1+\omega^2U_2$ can shift with $\omega$ and depends on the scalar chart. The quantum-period expansion at fixed parameters is generally nonuniform near a shrinking cycle, so its collision limit cannot be taken term by term, see~\cite{dunne2017mathieu,misumi2025degenerate}. Higher-order collisions and collisions with poles require their own local scaling. For analytic lift coefficients and fixed $\omega>0$, the exact monodromy matrix $M_{\det}$ remains analytic through a collision of classical turning points. Therefore, $\mathcal{D}_{\text{cl}}$ is generally distinct from the boundaries of the tongue, $\mathcal{F}=\pm 1$. All these facts are important for the construction of the semiclassical, i.e. leading-order monodromy via Stokes connection graph.

\subsection{Semiclassical monodromy and tongue boundaries}

Recall that for analytic coefficients the Stokes graph is formed by curves emanating from turning points $\tau_*$ and satisfying
\begin{equation}\label{eq:WKB-Stokes-graph}
    \im\left(e^{-i\theta}\int_{\tau_*}^{\tau}\lambda_0\right)=0, \quad \theta=\arg\omega.
\end{equation}
This graph organizes the domains where WKB solutions can be connected. A Stokes chamber is a region of parameters and summation direction where this connection pattern is fixed. We first take isolated simple turning points and paths, avoiding poles~\cite{iwaki2014clusters}.

We divide a path representing one forcing period into $N$ transport segments $\eta_j$ on the spectral cover. The leading exponential transport along the segment $j$ is
\begin{equation}\label{eq:WKB-leading-transport}
    T_j^{(0)}=\begin{pmatrix}e^{A_j/\omega}& 0\\ 0 & e^{-A_j/\omega}\end{pmatrix}, \quad A_j=\int_{\eta_j}\lambda_0,
\end{equation}
where $A_j$ is a segment action (do not be confused with the lift matrix $A(\tau)$). Let $C_j$ be the Airy connection or sheet-exchange matrix encountered after $\eta_j$, with consistent basis normalizations and orientations. In this WKB basis, the leading connection product is given by~\cite{sueishi2020quantization,sueishi2021anomaly}
\begin{equation}\label{eq:WKB-monodromy-product}
    \mathcal{P}^{(0)}=C_NT_N^{(0)}\cdots C_1T_1^{(0)}.
\end{equation}
We convert the initial and final WKB bases to the scalar coordinates, then restore eq.~\eqref{eq:WKB-gauge-monodromy}. This gives us a semiclassical (leading-order) approximation for $M_{\det}^{(0)}$. The condition $\tr M_{\det}^{(0)}=\pm 2$ estimates the boundaries of the tongue in this order. This construction cannot resolve exponentially narrow features with the required exponential accuracy. For example, the correction $\int\lambda_1$ already contributes at order one to the exponent, and a finite-order phase error can exceed an exponentially small bandwidth. Using the recurrence relation~\eqref{eq:WKB-momentum-recurrence}, we can obtain the all-orders WKB expansion. They retain the full formal expansions of actions and amplitudes, but typically do not converge. Moreover, they do not uniquely determine exponentially small contributions without additional analytic conditions. The exact WKB uses Borel resummation and Stokes connection formulas resolve these issues. For pedagogical introduction, see Part~I of~\cite{iwaki2025lectures} and Sec.~2 of~\cite{bucciotti2023anharmonic}.

\subsection{Borel resummation, DDP relations and exact monodromy}

As we have mentioned, formal WKB series are generally factorially divergent, Gevrey-I. Exact WKB uses Borel-Laplace resummation, which forms the Borel transform of the formal correction series and then takes its Laplace integral along a chosen direction. Under the corresponding analyticity and growth hypotheses, this constructs actual solutions with the prescribed asymptotic expansions~\cite{nikolaev2023wkb,iwaki2014clusters,nikolaev2024geometry}. The standard connection formulas in this approach assume simple turning points, admissible pole behavior, and a nonsingular summation direction. We should check these hypotheses in each scalar chart, with chart transitions included when needed. 

As usual, symbol $\mathcal{S}_\theta$ denotes the summation operation along the Borel-plane ray $e^{i\theta}\mathbb{R}_{>0}$, with the leading exponential factored out when summing a Voros symbol. In a nonsingular direction, we have $\mathcal{V}_{\gamma}^{(\theta)}=\mathcal{S}_\theta\!\left[\mathcal{V}_{\gamma}^{\text{form}}\right]$. A regular saddle trajectory joins two distinct simple turning points without passing through a pole, and its lift to the two sheets defines a closed saddle cycle $\gamma_0$. Across a singular direction with one such trajectory, the lateral sums obey the Delabaere-Dillinger-Pham (DDP) relation~\cite{delabaere1997semiclassical,dillinger1993resurgence,iwaki2014clusters},
\begin{equation}\label{eq:WKB-DDP-relation}
    \mathcal{S}_{\theta^-}\!\left[\mathcal{V}_{\gamma}^{\text{form}}\right]=\mathcal{S}_{\theta^+}\left[\mathcal{V}_{\gamma}^{\text{form}}\right]\left(1+\mathcal{S}_{\theta^+}\left[\mathcal{V}_{\gamma_0}^{\text{form}}\right]\right)^{(\gamma_0\circ\gamma)},
\end{equation}
where $\theta^-$ and $\theta^+$ denote the directions just below and above the singular Borel ray, and $\gamma_0\circ\gamma$ is the oriented intersection number. The saddle cycle is oriented as $\operatorname{Re}(e^{-i\theta}J_{\gamma_0})<0$, making its Voros exponential small. The sign in eq.~\eqref{eq:WKB-DDP-relation} uses $\theta=\arg\omega$. Using $\arg(1/\omega)$ reverses the lateral labels and the exponent sign. The jumps reconcile the connection formulas on the two sides, leaving the physical monodromy unchanged after the corresponding basis changes~\cite{sueishi2020quantization,sueishi2021anomaly}. They are distinct from the classical parameter monodromy in eq.~\eqref{eq:WKB-period-continuation}.

To construct exact monodromy, we should treat quantum transport, amplitudes, connection factors, and endpoint gages consistently. Let $W_\theta(\tau)$ be the invertible matrix whose columns are $(\psi_{\pm},\omega\psi_{\pm}')^{T}$ for two Borel-resummed WKB solutions, analytically continued along the forcing period using these connection formulas. Then, we have the following expression for the exact monodromy,
\begin{equation}\label{eq:WKB-exact-monodromy}
    M_{\det}^{\text{exact}}=L(2\pi)^{-1}W_\theta(2\pi)W_\theta(0)^{-1}L(0)=M_{\det}, \quad \mathcal{F}^{\text{exact}}=\frac{1}{2}\tr M_{\det}^{\text{exact}}.
\end{equation}
For a particular drive, an explicit tongue-boundary condition requires evaluating the transport and connection factors along the forcing cycle. Eq.~\eqref{eq:WKB-exact-monodromy} gives the reconstruction formula, while its evaluation depends on the chosen model and the Stokes chamber. If several scalar charts are used, their transition matrices are included in the continuation. Closed quantum periods alone need not determine every connection factor, since relative paths and pole data may also enter. The exact bandgap classification is eq.~\eqref{eq:WKB-Floquet-classification}, with the band-edge condition $\mathcal{F}^{\text{exact}}=\pm 1$. The exact monodromy describes the deterministic tongue structure, while the classical curve organizes its semiclassical scales. This structure provides the reference for the analysis of noise-induced phase slips and tongue melting.

To finalize, let us mention several examples of the WKB method in application to different systems. The Mathieu equation provides a standard example of the relation between WKB expansions, resurgence, and spectral bands, discussed in~\cite[Secs.~III and V]{dunne2017mathieu}. More examples of genus-1 systems are considered in~\cite{basar2017quantum,misumi2025degenerate}. The construction of quantum periods for higher-genus curves is developed in~\cite{fischbach2019wkb,ture2026quantumgeo}.

\section{Large deviations define second exponential scale}
\label{sec:LDP-exponential-scale}

Previously, we have discussed the melting of Arnold tongues and introduced the FW action. It was used to quantify phase slips, rotation number accuracy, and metastable locking. We now connect this stochastic variational problem to the deterministic WKB construction from sec.~\ref{sec:WKB-deterministic-skeleton}. In this section, we focus on how the optimal fluctuation cost depends on the slow frequency $\omega$ and when its exponential weight is comparable to a specified WKB-contribution.

We use fixed noise generators with constant channel intensities $D_r=D\nu_r$, where $D>0$ is the common noise scale and $\nu_r\geq0$ are fixed relative strengths. In slow time $\tau=\omega t$, we use $\varepsilon=D/\omega$ and $h_r(\phi)=\sqrt{\nu_r}\,V_r(\phi)$, where $V_r$ is the scalar coefficient of the phase generator. The local diffusion coefficient in physical time is $Da(\phi)$, with $a(\phi)=\sum_{r=1}^{3}h_r(\phi)^2=\sum_{r=1}^{3}\nu_rV_r(\phi)^2$, as in eqs.~\eqref{eq:weak-noise-scale} and \eqref{eq:fw-diffusion-coefficient}. For a measurable event $\mathcal{E}$ of continuously lifted phase paths over one forcing period, we fix the initial phase and set $\tau(0)=0$. Its physical-time cost is defined through the slow-time functional of eq.~\eqref{eq:fw-control-action}:
\begin{equation}\label{eq:LDP-event-action}
    \mathcal{A}_{\mathcal{E}}(\omega)=\omega\inf_{\gamma\in\mathcal{E}}\mathcal{I}_{[0,2\pi]}[\gamma],\qquad \frac{\mathcal{A}_{\mathcal{E}}(\omega)}{D}=\frac{\inf_{\gamma\in\mathcal{E}}\mathcal{I}_{[0,2\pi]}[\gamma]}{\varepsilon},
\end{equation}
where $\gamma$ denotes an absolutely continuous lifted path, parametrized by $\tau$. The cost is infinite for inaccessible paths and paths lacking absolute continuity. For fixed $\omega>0$, the small-noise path LDP~\cite{freidlin2012random} gives us
\begin{equation}\label{eq:LDP-event-probability}
    \lim_{D\to 0}-D\log\mathbb{P}_{D,\omega}(\mathcal{E})=\mathcal{A}_{\mathcal{E}}(\omega),
\end{equation}
provided $\mathcal{E}$ is a continuity set for the rate function, which means that its interior and closure have the same minimum action in the uniform path topology. Here, the probability $\mathbb{P}_{D,\omega}$ refers to the prescribed initial data. The event may depend on $\omega$, as for the moving-fold constraints below, although this limit holds with $\omega$ fixed. The same definition is valid for other specified observation intervals. For the unrestricted escape problem of eq.~\eqref{eq:fw-slip-barriers}, it instead gives $R_\pm^{\text{FW}}=\omega S_\pm^{\text{FW}}$, as in eq.~\eqref{eq:fw-time-normalization}. For a one-period event with fixed initial data, this barrier is not needed since it itself optimizes the departure and the transition time. Notice that we discuss here weak-noise LDP, which is different from long-time current LDP at fixed noise, used in~\cite{touchette2018markov,barato2018fluctuations}.

\subsection{Optimal fluctuations and random monodromy}

The physical-time action associated with eq.~\eqref{eq:fw-path-action} has the following FW Hamiltonian~\cite{grafke2019rare},
\begin{equation}\label{eq:LDP-Hamiltonian}
    H_{\text{FW}}(\phi,\tau,p,p_\tau)=\omega p_\tau+f(\phi,\tau)p+a(\phi)p^2,
\end{equation}
where $f$ is the deterministic phase drift, $p$ and $p_\tau$ are the momenta conjugate to $\phi$ and $\tau$, dots denote physical-time derivatives. The characteristic equations are
\begin{equation}\label{eq:LDP-Hamilton-equations}
\begin{aligned}
    \dot\phi&=f+2a(\phi)p, & \dot p &=-p\,\partial_{\phi} f-p^2\partial_{\phi} a(\phi),\\
    \dot\tau&=\omega, & \dot p_{\tau} &=-p\,\partial_{\tau} f.
\end{aligned}
\end{equation}
The branch $p=0$ reproduces a deterministic phase motion. We call \emph{stochastic instantons} the non-zero momentum characteristics that minimize the action subject to the event’s boundary conditions, while other characteristics are stationary paths. For a regular minimizer on $[t_0,t_1]$ with $a>0$, the action is
\begin{equation}\label{eq:LDP-action-on-instanton}
    \mathcal{A}_{\mathcal{E}}(\omega)=\int_{t_0}^{t_1}dt\,a(\phi)p^2,\quad p=\frac{\dot{\phi}-f}{2a(\phi)}.
\end{equation}
When $a$ vanishes, we should check accessibility through the control formulation in eq.~\eqref{eq:fw-control-action}. The deterministic WKB construction supplies the locked orbits, their separators, and the fold geometry that specify boundary data for these optimal paths. But it does not determine their costs without the noise fields.

We can use the same control principle for the M\"{o}bius monodromy. Let $A(\tau)$ and $K_r$ be the deterministic lift matrix and the fixed traceless noise matrices of eq.~\eqref{eq:lift-matrices}. Their projective vector fields are $f$ and $V_r$, respectively. For one period $T=2\pi/\omega$, we take square-integrable physical-time controls $\bm{v}\in L^2([0,T],\mathbb{R}^3)$, with cost $\int_0^Tdt\,\sum_r v_r(t)^2/4$, and set
\begin{equation}\label{eq:LDP-controlled-lift}
    \dot G_{\bm{v}}=\left[A(\tau)+\sum_r\sqrt{\nu_r}\,v_r(t)K_r\right]G_{\bm{v}},\,\, \dot\tau=\omega,\,\, M_\omega[\bm{v}]=G_{\bm{v}}(T),
\end{equation}
with $\tau(0)=0$ and $G_{\bm{v}}(0)=I$. For fixed $\omega$, the path LDP for the linear lift, followed by contraction under endpoint evaluation, gives us the rate function at speed $1/D$ for a target matrix $\mathcal{N}\in\mathrm{SL}(2,\mathbb{R})$~\cite{freidlin2012random,dembo2010deviations,touchette2009deviations},
\begin{equation}\label{eq:LDP-monodromy-contraction}
    \mathcal{I}_\omega(\mathcal{N})=\frac14\inf_{\bm{v}:\,M_\omega[\bm{v}]=\mathcal{N}}\int_0^Tdt\,\sum_r v_r(t)^2,\qquad \mathcal{I}_\omega(M_{\det})=0.
\end{equation}
Again, the infimum is $+\infty$ if the target is unreachable. For our smooth periodic coefficients, $M_{\det}=M_\omega[\bm{0}]$ is its unique zero, since controls whose costs tend to zero converge to the deterministic endpoint. In contrast to the action for a single phase path, this contraction preserves the individual matrix generators along with their common controls. Naturally, an endpoint matrix does not carry the complete integer winding of the continuously lifted phase. As we will see below, a change in monodromy and a completed phase slip must be specified as distinct events.

\subsection{How optimal action depends on forcing frequency}

\noindent\textbf{Activation away from folds.} Suppose that we have an admissible activated route which connects a frozen stable point $\phi_s(\tau)$ to a neighboring unstable point $\phi_u^{(j)}(\tau)$, with $a>0$ along the route. These points are roots of $f(\phi,\tau)=0$ with $\tau$ fixed and $j$ labels the exit direction on the phase lift. The zero-energy activation branch of the frozen Hamiltonian has $p=-f/a$ and $\dot\phi=-f$, so we obtain
\begin{equation}\label{eq:LDP-frozen-barrier}
    \Delta\mathcal{A}_j(\tau)=-\int_{\phi_s(\tau)}^{\phi_u^{(j)}(\tau)}d\phi\,\frac{f(\phi,\tau)}{a(\phi)}>0.
\end{equation}
If hyperbolicity and the barrier remain nonzero, the attracting orbit can reach the selected departure phase at zero cost, and the optimal activated segment is fast compared to $1/\omega$, the adiabatic estimate is~\cite{freidlin2012random,berglund2006noise} given by
\begin{equation}\label{eq:LDP-fast-action}
    \mathcal{A}_{\mathcal{E}}(\omega)=\mathcal{A}_0+o(1),\qquad \mathcal{A}_0=\min_{\text{admissible }(\tau,j)}\Delta\mathcal{A}_j(\tau)>0.
\end{equation}
This estimate implies that there is no cheaper nonadiabatic route, which realizes the event. It is not a replacement for the periodic-orbit minimization in eq.~\eqref{eq:fw-slip-barriers}, which allows nonadiabatic escape paths~\cite{dykman2005activated}. For autonomous additive-noise dynamics, it reduces to the potential-difference calculation that we have already illustrated by the Adler example, see.

\noindent\textbf{A rare deviation within a moving fold layer.} Then, we consider a nondegenerate fold of the real critical curve, corresponding to a simple turning point of the leading WKB problem~\cite{kristiansen2024turning}. At $(\phi_f,\tau_f)$, we have $f=\partial_{\phi} f=0$ and define
\begin{equation}\label{eq:LDP-turning-coefficients}
    \alpha_f=\left.\partial_\tau f\right|_f\neq0,\qquad \beta_f=\left.\frac12\partial_{\phi}^2f\right|_f\neq0,\qquad a_f=a(\phi_f)>0.
\end{equation}
With variables $x=\phi-\phi_f$ and $z=\tau-\tau_f$, we have the following leading local equation, obtained by freezing the diffusion coefficient at the fold,
\begin{equation}\label{eq:LDP-turning-normal-form}
    dx=(\alpha_f z+\beta_f x^2)\,dt+\sqrt{2Da_f}\,dW_t,\quad dz=\omega\,dt,
\end{equation}
where $W_t$ is a standard scalar Brownian motion representing the combined local noise channels. The rescaling $t-t_f=(|\alpha_f\beta_f|\omega)^{-1/3}s$, $x=\left[(|\alpha_f|\omega)/(|\beta_f|^2)\right]^{1/3}y$, $z=\omega(t-t_f)$ (where $t_f$ is the physical time at which $\tau=\tau_f$), gives us the normalized equation and its noise strength,
\begin{equation}\label{eq:LDP-turning-noise}
    dy=[\operatorname{sgn}(\alpha_f)s+\operatorname{sgn}(\beta_f)y^2]\,ds+\sqrt{2\varepsilon_f}\,dW_s,\qquad \varepsilon_f=\frac{Da_f|\beta_f|}{|\alpha_f|\omega}.
\end{equation}
\begin{figure}
    \centering
    \includegraphics[width=0.8\linewidth]{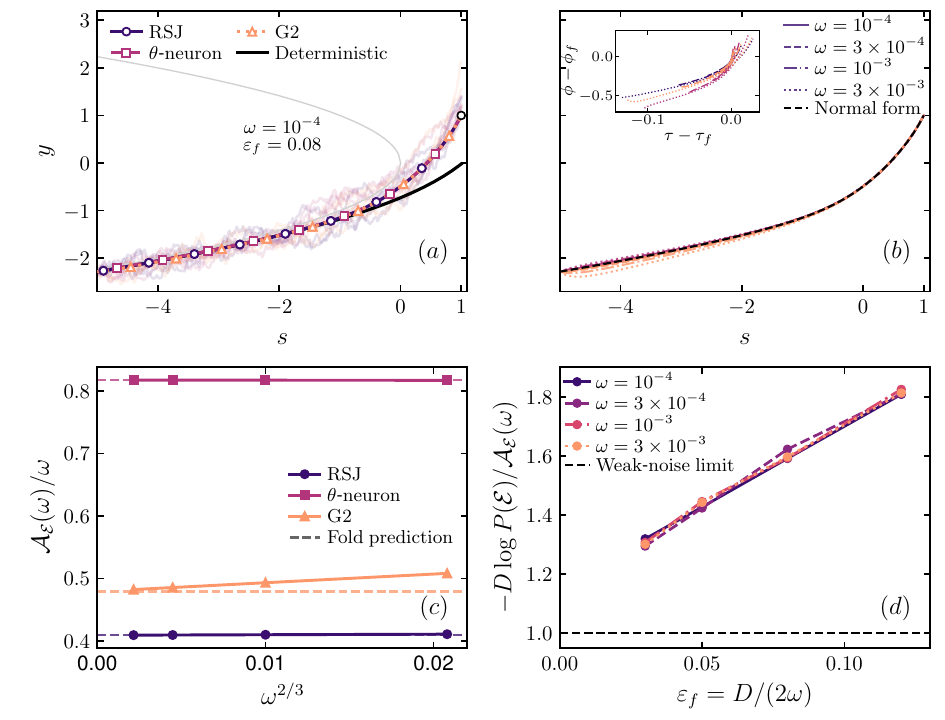}
    \caption{Rare fluctuations near a moving fold. (a) Action-minimizing paths for the RSJ, $\theta$-neuron, and G2 models in rescaled fold coordinates $(s,y)$ at $\omega=10^{-4}$ and $\varepsilon_f=0.08$. Thin colored curves show representative noisy realizations conditioned on $\mathcal{E}=\{y(1)\geq1\}$. The black curve shows the deterministic trajectory, and the gray curve shows the frozen repelling branch. (b) Optimal paths at different forcing frequencies approach the normal-form minimizer, shown by the black dashed curve. The inset displays the paths in the original coordinates relative to the fold. (c) Minimum physical-time action $\mathcal{A}_{\mathcal{E}}(\omega)$ divided by $\omega$, plotted against $\omega^{2/3}$. Dashed lines indicate the respective fold predictions. The limiting constants give $\mathcal{A}_{\mathcal{E}}(\omega)\sim C_{\mathcal{E},f}\omega$, with finite-frequency corrections most visible for G2. (d) Normalized logarithmic event probability for the RSJ model versus $D/(2\omega)$ at different forcing frequencies. For additive phase noise, $\varepsilon_f=D/[2A|\sin\tau_f|\omega]$, so the plotted variable equals $\varepsilon_f$ when $A|\sin\tau_f|=1$. The black dashed line marks the large-deviation prediction $-D\log P(\mathcal{E})/\mathcal{A}_{\mathcal{E}}(\omega)\to1$ as $D\to0$ at fixed $\omega$. Error bars represent $0.95$ confidence intervals}
    \label{fig:ldp-folds}
\end{figure}
Let $\widehat{\mathcal{E}}_f$ be the limiting event defined by the entrance and exit constraints in the rescaled variables. Its normalized minimum action is given by
\begin{equation}\label{eq:LDP-normalized-fold-action}
    \widehat{\mathcal{A}}_{\mathcal{E},f}=\frac14\inf_{y\in\widehat{\mathcal{E}}_f}\int_{s_{-}}^{s_{+}}ds\,\left[y'(s)-\operatorname{sgn}(\alpha_f)s-\operatorname{sgn}(\beta_f)y(s)^2\right]^2
\end{equation}
where the variable $y$ is absolutely continuous, $y'=dy/ds$, and $s_{-}$ and $s_{+}$ are the entry and exit times specified by the event, with any allowed exit time variation included in the infimum. This is the usual action for the normalized noisy moving saddle-node~\cite{herbert2017escape}. We assume that the rescaled minimum actions converge to $0<\widehat{\mathcal{A}}_{\mathcal{E},f}<\infty$, the local expansion is valid along the minimizing paths, and that their cost outside the layer is $o(\omega)$. If no cheaper route realizes the same event after rescaling of the action, we obtain
\begin{equation}\label{eq:LDP-turning-action}
    \boxed{\mathcal{A}_{\mathcal{E}}(\omega)=C_{\mathcal{E},f}\omega+o(\omega),\qquad C_{\mathcal{E},f}=\frac{|\alpha_f|}{a_f|\beta_f|}\widehat{\mathcal{A}}_{\mathcal{E},f}.}
\end{equation}
Its stochastic exponent is consequently $C_{\mathcal{E},f}\omega/D$ and the local weak-noise regime requires $\varepsilon_f\to 0$. Then, an ordinary deterministic fold passage has zero action, and a positive constant describes a prescribed rare deviation from that passage. The entrance and exit constraints are fixed in the limiting rescaled coordinates as $\omega\to 0$. Thresholds, which are fixed in the original coordinates, can define a different event. Moreover, a local fold deviation does not need produce a completed phase slip. For an annihilating fold, $\alpha_f\beta_f>0$, one example is the crossing a specified exit section before a fixed rescaled deadline earlier than its deterministic crossing time~\cite{herbert2017escape}. An additional activation barrier of order one changes the leading scaling, while $a_f=0$ requires a different local control analysis. For example, the forced $\theta$-neuron has a fold at $\phi_f=0$ modulo $2\pi$ and $B+A\cos\tau_f=0$. With additive phase noise, $a_f=1$, $\alpha_f=-2A\sin\tau_f$, and $\beta_f=1/2$, so $C_{\mathcal{E},f}=4|A\sin\tau_f|\widehat{\mathcal{A}}_{\mathcal{E},f}$ when $A\sin\tau_f\neq0$. The coefficient still requires the chosen event's normalized minimization. The moving-fold layer has a slow-phase width $O(\omega^{2/3})$. This differs from the noise-induced parameter rounding $\delta_{\text{round}}\propto D_{\text{eff}}^{2/3}$ in eq.~\eqref{eq:intrinsic-tongue-rounding}, which concerns a saddle-node of the return map at fixed $\omega$. These two scalings concern different local problems. Periodically driven escape can also exhibit intermediate regimes~\cite{ryvkine2004scaling}. Fig.~\ref{fig:ldp-folds} illustrates these predictions for the prescribed fold-layer event $\mathcal{E}=\{y(1)\geq 1\}$. The optimal paths for the three models approach the normalized fold minimizer as $\omega$ decreases, while conditioned noisy trajectories illustrate fluctuations around these paths, see fig.~\ref{fig:ldp-folds}(a,b). The convergence of $\mathcal{A}_{\mathcal{E}}(\omega)/\omega$ towards the corresponding fold predictions supports eq.~\eqref{eq:LDP-turning-action}, see fig.~\ref{fig:ldp-folds}(c). Fig.~\ref{fig:ldp-folds}(d) compares the event probabilities with eq.~\eqref{eq:LDP-event-probability}, showing the approach towards the logarithmic weak-noise prediction as $\varepsilon_f$ decreases, with appreciable finite-noise corrections over the displayed range.

\noindent\textbf{A sustained departure from the drift.} Finally, suppose that the observation interval is asymptotic to a fixed positive multiple of $1/\omega$ and that the minimum mean control cost over this interval converges to a finite positive value. Then the total action satisfies $\mathcal{A}_{\mathcal{E}}\sim c_{\mathcal{E}}/\omega$ with $c_{\mathcal{E}}>0$. The corresponding exponent is $c_{\mathcal{E}}/(D\omega)$. Constructing one such controlled path gives us only an upper bound. The stated scaling requires a matching lower estimate that excludes cheaper localized paths.

\subsection{Comparison of exponential scales}

Now we can compare the decay exponents supplied by the deterministic WKB construction and the stochastic variational problem. We equate them and this defines a formal matching scale. A crossover in an observable additionally requires both contributions to enter that observable with the same normalization and to dominate its remaining corrections. For an active WKB contribution and a specified rare event $\mathcal{E}$, we write the corresponding weights as follows,
\begin{equation}\label{eq:LDP-general-WKB-weight}
    w^{\text{WKB}}_{\gamma}=C^{\text{WKB}}_{\gamma}\exp[-\mathcal{S}^{\text{WKB}}_{\gamma}(\omega)/\omega],\quad w^{\text{FW}}_{\mathcal{E}}=C^{\text{FW}}_{\mathcal{E}}\exp[-\mathcal{A}_{\mathcal{E}}(\omega)/D].
\end{equation}
In the eq.~\eqref{eq:LDP-general-WKB-weight}, $C^{\text{WKB}}_{\gamma}$ and $C^{\text{FW}}_{\mathcal{E}}$ are prefactors that can both depend on $\omega$ and the prefactor $C^{\text{FW}}_{\mathcal{E}}$ also depends on $D$. The index $\gamma$ labels the spectral cycle or a combination of relative paths that contribute to the chosen observable, while the positive quantity $\mathcal{S}^{\text{WKB}}_{\gamma}$ corresponds to the decay action, which is extracted from the active combination of spectral-curve periods and relative path data in deterministic monodromy. The normalization of $\mathcal{S}^{\text{WKB}}_{\gamma}$ normalization belongs to the observable, i.e. a tunneling amplitude, a probability, and a band width do not need to use the same fraction of a cycle action. In contrast, the FW cost is a real optimal-control action and depends on the noise channels. We match these two exponents, which defines the crossover scale,
\begin{equation}\label{eq:LDP-general-crossover}
    \boxed{\frac{\mathcal{S}^{\text{WKB}}_{\gamma}(\omega)}{\omega}\sim\frac{\mathcal{A}_{\mathcal{E}}(\omega)}{D},\qquad D_{\text{match}}(\omega)\sim\omega\frac{\mathcal{A}_{\mathcal{E}}(\omega)}{\mathcal{S}^{\text{WKB}}_{\gamma}(\omega)}.}
\end{equation}
For a non-degenerate WKB action, $\mathcal{S}^{\text{WKB}}_{\gamma}(\omega)=S_0>0$, the three scenarios discussed above give us the following formal matching scales,
\begin{equation}\label{eq:LDP-matching-regimes}
\begin{array}{c|c|c}
    \text{optimal fluctuation} & \mathcal{A}_{\mathcal{E}}(\omega) & D_{\text{match}}(\omega)\\ \hline
    \text{adiabatic activation} & \mathcal{A}_0+o(1) & (\mathcal{A}_0/S_0)\omega\\
    \text{rare fold-layer deviation} & C_{\mathcal{E},f}\omega+o(\omega) & (C_{\mathcal{E},f}/S_0)\omega^2\\
    \text{sustained departure} & c_{\mathcal{E}}/\omega\,[1+o(1)] & c_{\mathcal{E}}/S_0
\end{array}
\end{equation}
These scales compare selected exponential weights and their interpretation as an observable crossover requires a common observable expansion with controlled remainders (we will illustrate this statement by canards persistence in the next section). In the last row, the formal matching scale does not tend to zero, so it lies outside a joint limit with $D\to 0$. More generally, if $\mathcal{A}_{\mathcal{E}}\sim c_A\omega^\mu$ and $\mathcal{S}^{\text{WKB}}_{\gamma}\sim c_S\omega^\nu$, with positive coefficients, we have the matching scale
\begin{equation}\label{eq:LDP-general-scaling-law}
    D_{\text{match}}(\omega)\sim\frac{c_A}{c_S}\omega^{1+\mu-\nu}.
\end{equation}
To compare two exponentially small weights, we require $\nu<1$, and a matching scale tending to zero additionally requires $1+\mu-\nu>0$. Near a shrinking cycle, its continued action and the relevant local approximation must be determined before applying this formula.

The eq.~\eqref{eq:LDP-matching-regimes} provides the conditional joint limit estimates. For fixed $\omega$, the respective LDP in eq.~\eqref{eq:LDP-event-probability} alone does not establish them: the minimizing paths and logarithmic remainders must be controlled along the chosen $D(\omega)$. For $S_0>0$, the required errors and logarithms of prefactors are $o(1/\omega)$. For the specified rare fold-layer event, $D_{\text{match}}\asymp\omega^2$ equates its stochastic decay exponent with a non-degenerate WKB decay exponent. This matching does not establish a threshold for tongue melting. Our metastable locking uses the completed-slip criterion of eq.~\eqref{eq:metastable-locking-condition}.

The leading moving-fold model provides a concrete local connection between the two constructions. Consider the orientation $\alpha_f>0$ and $\beta_f>0$, for which a square-integrable scalar control $v(s)$ gives
\begin{equation}\label{eq:LDP-controlled-fold}
    y'=s+y^2+v(s),\qquad \widehat{\mathcal{A}}_{\mathcal{E},f}=\frac14\inf_{v:\,y\in\widehat{\mathcal{E}}_f}\int_{s_{-}}^{s_{+}}ds\,v(s)^2,
\end{equation}
with the prescribed entrance data. On an interval where the Riccati coordinate remains finite, set $y=-u'/u$ with $u(s_{-})=1$, so that $u>0$ throughout this interval. Then, we have $u''+[s+v(s)]u=0$. At $v=0$, this is the Airy equation describing the deterministic simple turning point. The rare-event constraints become conditions on its logarithmic derivative. For example, $y(1)\geq1$ becomes $u'(1)+u(1)\leq0$. The normalized FW action therefore measures the least quadratic control needed to deform this local Airy problem so that the prescribed event occurs. The physical action follows from eq.~\eqref{eq:LDP-turning-action}, including its dependence on the diffusion coefficient $a_f$. This local correspondence relates deterministic turning-point geometry to stochastic optimal control without identifying their actions or promoting the formal matching scales to observable crossover thresholds.

\section{Canards and their persistence}
\label{sec:canard-persistence}

As mentioned earlier, a canard follows an attracting slow branch and then remains close to a repelling one for an $O(1)$ interval of slow time. We now ask whether this repelling passage survives with noise, motivated by~\cite{sowers2008canards,berglund2012hunting}. The main mechanism is simple: a fluctuation acquired near the entrance is amplified exponentially along the repelling branch. This makes canard shadowing more restrictive than small local noise. Avoiding a completed phase slip over many periods is a further requirement, since an early departure from a canard need not change the lifted winding.

\subsection{Deterministic selection and accumulated repulsion}

First, let us recall which deterministic trajectories are being perturbed. We consider the Riccati equation for variable $\Phi$, $\omega\Phi'=q_0+q_1\Phi+q_2\Phi^2$ and the Riccati discriminant determines the local contraction and expansion rates on the attracting $\mathcal{C}_0^{a}$ and repelling $\mathcal{C}_0^{r}$ critical branches are determined by
\begin{equation}\label{eq:canard-critical-discriminant}
    \Delta_0(\tau)=q_1(\tau)^2-4q_0(\tau)q_2(\tau),\quad
    \left.\partial_\phi f\right|_{\mathcal{C}_0^{a}}=-\sqrt{\Delta_0},\quad
    \left.\partial_\phi f\right|_{\mathcal{C}_0^{r}}=+\sqrt{\Delta_0},
\end{equation}
and these branches exist for $\Delta_0>0$. In a regular scalar reduction, simple real zeros correspond to nondegenerate folds and leading WKB turning points. A sequence of slow arcs, folds, and fast transitions defines an admissible canard itinerary only if the global return selects the required repelling passage~\cite{guckenheimer2001staircase,schurov2010torus}. For example, at a simple tongue edge, a nonidentity parabolic return map has a neutral periodic orbit. If its slow itinerary is a canard, the exact finite-$\omega$ representative satisfies
\begin{equation}\label{eq:canard-balanced-location}
    P_{\omega}'(\phi_{\text{can}}^\omega(0))
    =\exp\left[\frac1\omega\int_0^{2\pi}d\tau\,\partial_\phi f(\phi_{\text{can}}^\omega(\tau),\tau)\right]=1.
\end{equation}
This compensation of contraction and expansion, including transitions, defines a \emph{balanced canard}. So, neutrality alone does not establish the canard itinerary. The neighboring hyperbolic tongue can contain attracting and repelling canard cycles.
\begin{figure}[h!]
    \centering
    \includegraphics[width=\linewidth]{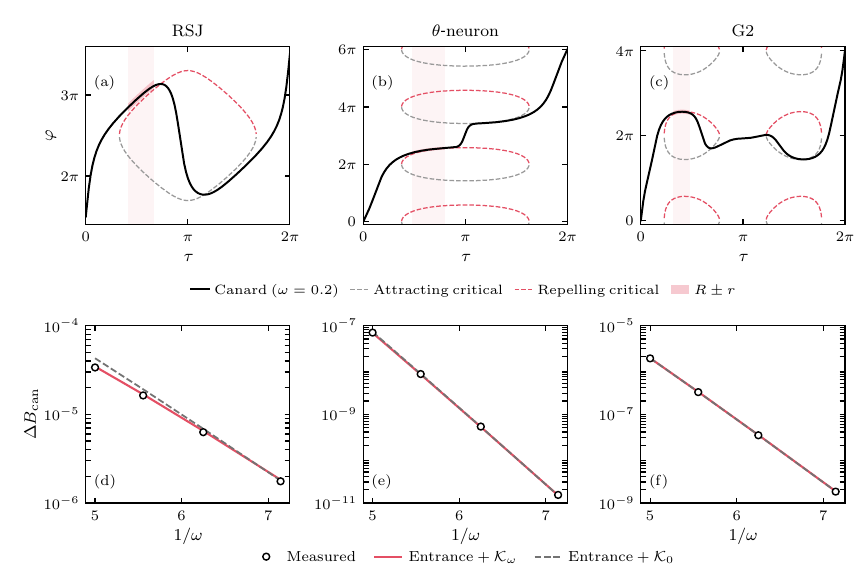}
    \caption{(a)-(c) Deterministic canards for RSJ, $\theta$-neuron, and G2 at $A=1.2$, $\omega=0.2$. Black curves show lifted phase over one period, gray and red dashed curves represent attracting and repelling critical branches. Pale shading marks selected repelling intervals, with darker tubes $\phi_r^\omega(\tau)\pm r$ around the finite-frequency reference, $r=0.15$. (d)-(f) Canard-window widths $\Delta B_{\text{can}}$ on logarithmic vertical axes versus $1/\omega$. Circles lengths of $B$ intervals, where the selected trajectory remains within the prescribed tube, solid red line represent the computed periodic entrance map and $\mathcal{K}_{\omega}$, gray dashed line use the same map and $\mathcal{K}_0$. The approximately linear trends support exponential narrowing over the displayed range.}
    \label{fig:deterministic-canards}
\end{figure}
We choose a compact repelling segment $[\tau_{\text{in}},\tau_{\text{out}}]$ away from folds and an exact solution $\phi_r^\omega$ following it. We introduce the following quantity, so-called \emph{canard budget},
\begin{equation}\label{eq:canard-growth-budget}
    \kappa_{\omega}(\tau)=\partial_\phi f(\phi_r^\omega(\tau),\tau)>0,\quad
    \mathcal{K}_{\omega}(\tau)=\int_{\tau_{\text{in}}}^{\tau}du\,\kappa_{\omega}(u).
\end{equation}
Our reference solves $\omega(\phi_r^\omega)'=f(\phi_r^\omega,\tau)$ and its $O(\omega)$ displacement from the critical curve exceeds the relevant entrance tolerance. Normal hyperbolicity gives us $\kappa_{\omega}=\sqrt{\Delta_0}+O(\omega)$~\cite{fenichel1979geometric,kuehn2015timescales} with leading action
\begin{equation}\label{eq:canard-growth-Abelian-integral}
    \mathcal{K}_0(\tau)=\int_{\tau_{\text{in}}}^{\tau}du\,\sqrt{\Delta_0(u)}.
\end{equation}
In a regular scalar chart, $\sqrt{\Delta_0}/2$ describes the leading WKB growth rate~\cite{kristiansen2024turning}. This means that $\mathcal{K}_0$ is twice the corresponding growth integral, not necessarily a full compact-cycle action. In linear order, an entrance displacement grows as
\begin{equation}\label{eq:canard-spends-budget}
    d(\tau)=d_{\text{in}}e^{\mathcal{K}_{\omega}(\tau)/\omega}.
\end{equation}
We write $\mathcal{K}_{\text{out}}=\mathcal{K}_{\omega}(\tau_{\text{out}})$ and want to track the reference within the radius $r$. It requires the entrance accuracy of order $re^{-\mathcal{K}_{\text{out}}/\omega}$. Suppose that WKB matching relates the entrance displacement to a control parameter $s$ (for instance, $B$) by the following law,
\begin{equation}\label{eq:canard-matching-coefficient}
    |d_{\text{in}}(s,\omega)|\sim|c_{\text{in}}(\omega)|\,|s-s_*|^\chi,\quad \chi>0,
\end{equation}
then the detuning scale is given by
\begin{equation}\label{eq:canard-width-from-entrance}
    \delta s_{\text{can}}\asymp
    \left(\frac{r}{|c_{\text{in}}(\omega)|}\right)^{1/\chi}
    e^{-\mathcal{K}_{\text{out}}/(\chi\omega)}.
\end{equation}
This scale is exponentially small when the action has a positive limit and the quantities $r$, $c_{\text{in}}$ are subexponential in $1/\omega$. The regular transverse matching has $\chi=1$ and several repelling passages require all entrance conditions. A WKB connection term alone does not determine a parameter width without this matching law. The examples of canard trajectories for our illustrative models are shown at fig.~\ref{fig:deterministic-canards}(a-f).

\subsection{Noise amplification}

Again, we work with fixed intensities of noise generators,
\begin{equation}\label{eq:canard-noise-fields}
    D_r=D\nu_r,\quad h_r(\phi)=\sqrt{\nu_r}\,V_r(\phi),\quad
    a(\phi)=\sum_{r=1}^{3}h_r(\phi)^2,\quad \nu_r\geq 0.
\end{equation}
In slow time, our stochastic equation is
\begin{equation}\label{eq:canard-slow-sde}
    d\phi_{\tau}=\frac{f(\phi_{\tau},\tau)}{\omega}\,d\tau+\sqrt{\frac{2D}{\omega}}\sum_{r=1}^{3}h_r(\phi_{\tau})\circ dW_{\tau}^r.
\end{equation}
We linearize the drift term and evaluate diffusion on the repelling reference,
\begin{equation}\label{eq:canard-linear-noisy-repeller}
    d\xi_{\tau}=\frac{\kappa_{\omega}(\tau)}{\omega}\xi_{\tau}\,d\tau+\sqrt{\frac{2D}{\omega}}\sum_rh_r(\phi_r^\omega(\tau))\,dW_{\tau}^r.
\end{equation}
For deterministic entry $d_{\text{in}}$, this Gaussian approximation gives us
\begin{equation}\label{eq:canard-linear-noise-solution}
\begin{gathered}
    \xi_{\tau}=e^{\mathcal{K}_{\omega}(\tau)/\omega}(d_{\text{in}}+Z_{\tau}),\quad
    \operatorname{Var}Z_{\tau}=D\mathcal{L}(\tau,\omega),\\
    \mathcal{L}(\tau,\omega)=\frac{2}{\omega}\int_{\tau_{\text{in}}}^{\tau}du\,a(\phi_r^\omega(u))e^{-2\mathcal{K}_{\omega}(u)/\omega}.
\end{gathered}
\end{equation}
The martingale $Z$ is the noise referred back to the entrance. With $\mathcal{L}_{\text{out}}=\mathcal{L}(\tau_{\text{out}},\omega)>0$, its characteristic scale is defined by
\begin{equation}\label{eq:canard-pathwise-control}
    \boxed{D_{\mathrm{path}}=\frac{r^2}{\mathcal{L}_{\text{out}}}e^{-2\mathcal{K}_{\text{out}}/\omega},\quad
    \frac{\operatorname{Var}\xi_{\text{out}}}{r^2}=\frac{D}{D_{\mathrm{path}}}.}
\end{equation}
Here, the exponent is doubled because the variance is the squared amplitude. For a regular entrance with uniformly positive diffusion and smooth, uniformly controlled coefficients, we can write down the $\mathcal{L}_{\text{out}}$ as follows,
\begin{equation}\label{eq:canard-regular-entrance-covariance}
    \mathcal{L}_{\text{out}}\sim\frac{a(\phi_r^\omega(\tau_{\text{in}}))}{\kappa_{\omega}(\tau_{\text{in}})}.
\end{equation}
For example, our forced $\theta$-neuron on a repelling subarc has $I(\tau)=B+A\cos\tau<0$, so we have
\begin{equation}\label{eq:canard-theta-example}
    \mathcal{K}_0(\tau_{\text{out}})=2\int_{\tau_{\text{in}}}^{\tau_{\text{out}}}du\,\sqrt{-B-A\cos u}.
\end{equation}
For additive noise, $a=1$, which gives the leading exponent $-4\int_{\tau_{\text{in}}}^{\tau_{\text{out}}}du\,\sqrt{-I(u)}/\omega$ in $D_{\mathrm{path}}$. So, $D/\omega\to0$ alone is insufficient. In the Gaussian model, $D/D_{\mathrm{path}}\to\infty$ makes even terminal tube inclusion improbable.

For globally active noise, we extend the exact canard back to $\tau_0<\tau_{\text{in}}$, writing $\phi_{\text{can}}^\omega=\phi_r^\omega$ on the selected segment. We define the signed growth and propagator
\begin{equation}\label{eq:canard-global-propagator}
    \lambda_{\omega}(\tau)=\partial_\phi f(\phi_{\text{can}}^\omega(\tau),\tau),\quad
    U_{\omega}(\tau,s)=\exp\left[\frac1\omega\int_s^\tau du\,\lambda_{\omega}(u)\right].
\end{equation}
Starting on the reference at $\tau_0$, the Gaussian approximation tells us that the entrance variance is $D\mathcal{L}_{\text{in}}$ with
\begin{equation}\label{eq:canard-global-entrance-variance}
    \mathcal{L}_{\text{in}}=\frac{2}{\omega}\int_{\tau_0}^{\tau_{\text{in}}}ds\,a(\phi_{\text{can}}^\omega(s))U_{\omega}(\tau_{\text{in}},s)^2.
\end{equation}
The independence of future Brownian increments gives the terminal variance and its characteristic Gaussian scale,
\begin{equation}\label{eq:canard-global-noise-scale}
    \operatorname{Var}\xi_{\text{out}}=D e^{2\mathcal{K}_{\text{out}}/\omega}(\mathcal{L}_{\text{in}}+\mathcal{L}_{\text{out}}),
    \boxed{D_{\mathrm{path}}^{\mathrm{glob}}=\frac{r^2 e^{-2\mathcal{K}_{\text{out}}/\omega}}{\mathcal{L}_{\text{in}}+\mathcal{L}_{\text{out}}}.}
\end{equation}
Attraction damps old errors, but continuously injected noise leaves fluctuations, typically of the order $\sqrt{D}$ away from folds~\cite{berglund2003geometric}. Earlier repulsion can instead make $\mathcal{L}_{\text{in}}$ exponentially large. An omitted repelling portion immediately preceding the selected segment, with action $\mathcal{K}_{\text{pre}}$ and nondegenerate noise near its beginning, can change the leading scale to $e^{-2(\mathcal{K}_{\text{pre}}+\mathcal{K}_{\text{out}})/\omega}$ up to subexponential factors. The compact-segment exponent is retained only when the total variance factor is subexponential. These Gaussian formulas identify fluctuation scales, not a nonlinear entrance law or survival probability through preceding folds and transitions.

\subsection{Canard persistence theorem}

Now we are ready to formulate the canard persistence theorem. We use lifted phases and $x_{\tau}=\phi_{\tau}-\phi_r^\omega(\tau)$, where $r$ is smaller than a fixed local phase neighborhood. Following the sample-path approach from~\cite{berglund2003geometric,berglund2006noise}, we formulate the following theorem,

\begin{theorem}[Persistence of a prescribed canard segment]\label{thm:canard-persistence}

Let $\omega>0$, $D>0$ and let $\phi_r^\omega$ solve the deterministic equation on $[\tau_{\text{in}},\tau_{\text{out}}]$, with $\kappa_{\omega}\geq\kappa_*>0$. Assume smooth coefficients and, in $|x|\leq r$,
\begin{equation}\label{eq:canard-tube-assumptions}
\begin{gathered}
    |f(\phi_r^\omega+x,\tau)-f(\phi_r^\omega,\tau)-\kappa_{\omega} x|\leq C_f x^2,\\
    a(\phi_r^\omega+x)\leq a_*,\quad |c(\phi_r^\omega+x)|\leq c_*,\quad
    c(\phi)=\sum_r h_r(\phi)h_r'(\phi)=\tfrac12a'(\phi),
\end{gathered}
\end{equation}
where $C_f,c_*\geq0$, $a_*>0$, and primes denote phase derivatives. Set
\begin{equation}\label{eq:canard-theorem-scales}
    \delta_{\omega}=\frac{r e^{-\mathcal{K}_{\text{out}}/\omega}}{1+C_fr/\kappa_*},\quad
    \overline{\mathcal{L}}_{\omega}=\frac{2a_*}{\omega}\int_{\tau_{\text{in}}}^{\tau_{\text{out}}}du\,e^{-2\mathcal{K}_{\omega}(u)/\omega}.
\end{equation}
If the entrance satisfies
\begin{equation}\label{eq:canard-deterministic-tube-margin}
    |d_{\text{in}}|+\frac{Dc_*}{\kappa_*}\leq(1-\eta)\delta_{\omega},
\end{equation}
for some $0<\eta<1$, then
\begin{equation}\label{eq:canard-local-pathwise-classification}
    \boxed{\mathbb{P}\left\{\sup_{\tau_{\text{in}}\leq\tau\leq\tau_{\text{out}}}|x_{\tau}|\geq r\right\}
    \leq2\exp\left[-\frac{\eta^2\delta_{\omega}^2}{2D\overline{\mathcal{L}}_{\omega}}\right].}
\end{equation}
The bound also holds conditionally on the entrance history whenever a random entrance satisfies eq.~\eqref{eq:canard-deterministic-tube-margin}. No positive lower bound on $a$ is required.
\end{theorem}

\begin{proof}
Set $G(\tau)=e^{\mathcal{K}_{\omega}(\tau)/\omega}$ and let $\sigma$ be the first tube exit. In It\^{o} form, before exit, we can write
\begin{equation}\label{eq:canard-exact-displacement}
    dx_{\tau}=\frac{\kappa_{\omega} x_{\tau}+R(x_{\tau},\tau)+Dc(\phi_{\tau})}{\omega}\,d\tau
    +\sqrt{\frac{2D}{\omega}}\sum_r h_r(\phi_{\tau})\,dW_{\tau}^r,
    \quad |R(x,\tau)|\leq C_fx^2.
\end{equation}
Multiplication by $G^{-1}$ removes linear growth. The stopped martingale satisfies
\begin{equation}\label{eq:canard-stopped-martingale}
    M_{\tau}=\sqrt{\frac{2D}{\omega}}\sum_r\int_{\tau_{\text{in}}}^{\tau\wedge\sigma}dW_u^r\,G(u)^{-1}h_r(\phi_u),\quad
    \langle M\rangle_{\tau_{\text{out}}}\leq D\overline{\mathcal{L}}_{\omega},
\end{equation}
where $\langle M\rangle$ denotes quadratic variation. Since $\omega^{-1}\int_{\tau_{\text{in}}}^{\tau_{\text{out}}}du\,G(u)^{-1}\leq1/\kappa_*$, the variables
\begin{equation*}
    Y_{\tau}=G(\tau)^{-1}x_{\tau},\quad
    H=|d_{\text{in}}|+\frac{Dc_*}{\kappa_*}+\sup_{\tau_{\text{in}}\leq u\leq\tau_{\text{out}}}|M_u|
\end{equation*}
obey $|Y_{\tau}|\leq H+(C_f/\omega)\int_{\tau_{\text{in}}}^{\tau}du\,G(u)|Y_u|^2$ up to exit. Quadratic integral comparison and $G'\geq\kappa_*G/\omega$ give us
\begin{equation}\label{eq:canard-nonlinear-comparison}
    |x_{\tau}|\leq\frac{G(\tau)H}{1-(C_f/\kappa_*)H[G(\tau)-1]},
\end{equation}
while the denominator is positive. On $\sup|M|<\eta\delta_{\omega}$, the entrance condition gives $H<\delta_{\omega}$, making the denominator positive and the bound smaller than $r$ throughout. Continuity excludes exit. The exponential martingale inequality now yields
\begin{equation*}
    \mathbb{P}\{\sup|M|\geq\eta\delta_{\omega}\}
    \leq2\exp\left[-\frac{\eta^2\delta_{\omega}^2}{2D\overline{\mathcal{L}}_{\omega}}\right].
\end{equation*}
The argument also holds conditionally on the entrance history.
\end{proof}

\begin{figure}[h!]
    \centering
    \includegraphics[width=0.95\linewidth]{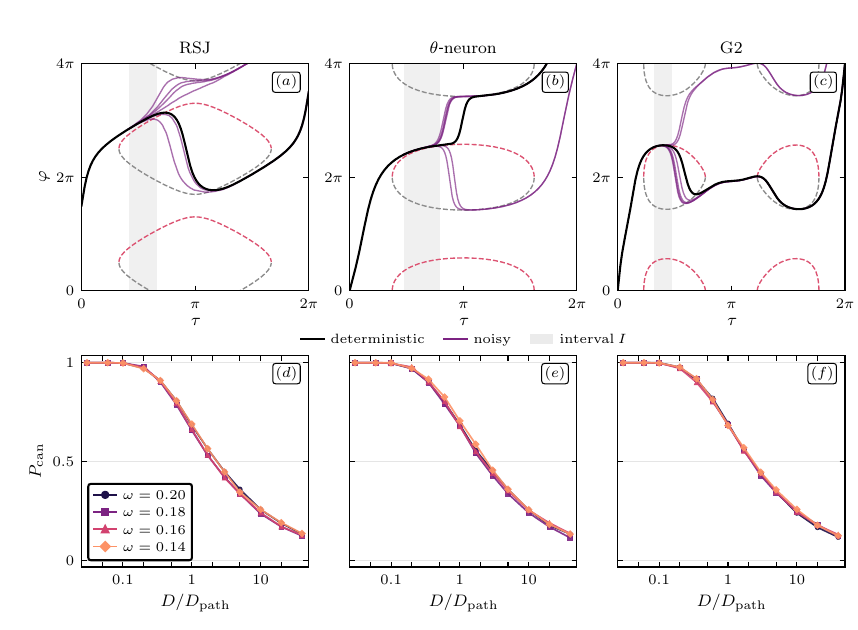}
    \caption{(a)-(c) Deterministic canards (black) and six trajectories with additive noise at $\omega=0.2$, $D=10D_{\mathrm{path}}$, for RSJ, $\theta$-neuron, and G2. Noise starts at the entrance to the shaded interval $I$, with every realization on its deterministic reference. (d)-(f) Local survival $P_{\mathrm{can}}=\mathbb{P}\{\sup_{\tau\in I}|\varphi_D-\varphi_{\mathrm{can}}^\omega|<r\}$ versus $D/D_{\mathrm{path}}$, with $r=0.15$ and respective intervals $[1.3,2.1]$, $[1.5,2.5]$, $[1.0,1.5]$. Each point uses $2000$ realizations. $B(\omega)$ retains a fixed fractional position in each measured canard window. This protocol isolates local amplification and does not measure the globally noisy entrance law.}
    \label{fig:noisy-canards}
\end{figure}

At the fig.~\ref{fig:noisy-canards}, we illustrate the proposed local noise-amplification mechanism. Fig.~\ref{fig:noisy-canards}(a)-(c) show individual noisy trajectories with deterministic trajectory, while fig.~\ref{fig:noisy-canards}(d)-(f) show the survival probability as a function of $D/D_{\mathrm{path}}$, with noise switched on at the entrance to the prescribed repelling segment. For uniformly controlled coefficients, a regular entrance, and a slow-time interval bounded below, the quantity $\overline{\mathcal{L}}_{\omega}$ is of the order of one. If the entrance margin is satisfied, nonlinear persistence occurs when
\begin{equation}\label{eq:canard-pathwise-threshold}
    D\ll\frac{\delta_{\omega}^2}{\overline{\mathcal{L}}_{\omega}}
    =\frac{r^2 e^{-2\mathcal{K}_{\text{out}}/\omega}}{(1+C_fr/\kappa_*)^2\overline{\mathcal{L}}_{\omega}}.
\end{equation}
This sufficient bound allows $r$ to be fixed and small. Its prefactor is conservative, so a sharp failure threshoild requires lower fluctuation estimates. The vanishing diffusion on the orbit can change the actual scale.

For globally active noise, let $X_{\text{in}}=\phi_{\tau_{\text{in}}}-\phi_r^\omega(\tau_{\text{in}})$. Conditioning gives us the unconditional consequence
\begin{equation}\label{eq:canard-global-persistence-bound}
\begin{gathered}
    \beta_{\text{in}}=\mathbb{P}\left\{|X_{\text{in}}|+\frac{Dc_*}{\kappa_*}>(1-\eta)\delta_{\omega}\right\},\\
    \boxed{\mathbb{P}\left\{\sup_{\tau_{\text{in}}\leq\tau\leq\tau_{\text{out}}}|x_{\tau}|\geq r\right\}
    \leq\beta_{\text{in}}+2e^{-\eta^2\delta_{\omega}^2/(2D\overline{\mathcal{L}}_{\omega})}.}
\end{gathered}
\end{equation}
Thus, switching noise off before entry is not necessary, but $\beta_{\text{in}}$ must be controlled. Full-itinerary survival additionally requires preceding tube-exit estimates. The fold coordinates of sec.~\ref{sec:LDP-exponential-scale} do not by themselves provide these estimates or the nonlinear entrance law. At fixed $\omega>0$ and finite observation time, small-noise convergence ensures shadowing from the deterministic reference as $D\to0$. The joint slow-forcing limit requires the stronger entrance and amplification controls above.

\subsection{Several passages and metastable locking}

We can extend the obtained result for several repelling passages. Suppose that an itinerary contains $m$ repelling segments per period. Then, we can apply the theorem with a common margin $\eta$ and set
\begin{equation}\label{eq:canard-uniform-variance-ratio}
    \mathfrak R=\max_j\frac{D\overline{\mathcal{L}}_{\omega,j}}{\delta_{\omega,j}^2},
\end{equation}
where $j$ includes all $N$ observed periods if bounds vary. Let $\beta_N$ bound each conditional entrance-failure probability, given successful earlier entrances and passages, including intervening attraction, folds, and transitions. Let $p_{\text{slip},N}$ denote the probability of at least one completed slip. The probability $P_{\mathrm{pers}}^{(N)}$ of tracking all prescribed repelling segments without a completed slip satisfies
\begin{equation}\label{eq:canard-persistence-regime}
    1-P_{\mathrm{pers}}^{(N)}\leq
    mN\left[\beta_N+2e^{-\eta^2/(2\mathfrak R)}\right]+p_{\text{slip},N}.
\end{equation}
Indeed, the first failed passage is either a bad entrance or an exit after good entry. Summing these conditional bounds and the slip probability requires no independence. For fixed $m$, a sufficient joint-limit criterion is
\begin{equation}\label{eq:canard-uniform-persistence-condition}
    \boxed{\mathfrak R\log(N+1)\longrightarrow0,\quad
    N\beta_N\longrightarrow0,\quad p_{\text{slip},N}\longrightarrow0.}
\end{equation}
Successful passage does not guarantee exponentially accurate re-entry. An attracting return can suppress accumulated errors, whereas a balanced edge supplies no strict linear contraction. Inside a hyperbolic tongue, a metastable reduction uniform on the observation scale~\cite{dykman2005activated,berglund2016phaseslips} gives the rates per forcing period
\begin{equation}\label{eq:canard-phase-slip-probabilities}
    p_{\text{slip},N}\simeq1-e^{-N\Gamma_\Sigma},\quad
    \Gamma_\Sigma=\Gamma_{+}+\Gamma_{-},\quad
    \Gamma_\pm=C_\pm e^{-R_\pm^{\text{FW}}/D},\quad
    R_\pm^{\text{FW}}=\omega S_\pm^{\text{FW}}.
\end{equation}
Then $N\Gamma_\Sigma\to0$ controls slips, provided actions, prefactors, and return estimates remain uniform as $\omega$ or tongue-edge distance varies. The matching of selected exponential weights in sec.~\ref{sec:LDP-exponential-scale} does not establish canard persistence. Persistence requires entrance accuracy, weak amplified noise, and rare completed slips over the observation time. Uniformly accessible slips at fixed positive noise preclude infinite no-slip survival. Our claim is finite-time canard shadowing together with metastable locking.

\section{Discussion}
\label{sec:mobius-discussion}

In this paper, we have considered the deterministic phase-locking phenomenon under the influence of stochastic perturbations that preserve the M\"{o}bius structure. In the limit of slow frequency of external drive, we have shown that WKB method can be applied to construct phase-locking domains in parameter space, i.e. Arnold tongues. This fact just reflects the Riccati-Schr\"{o}dinger correspondence that identifies the classical WKB curve of the genus $g\geq 1$ with the complexified slow curve. WKB action allows to reconstruct boundaries of tongues in the leading $\omega$-order, while the exact WKB method can resolve exponentially small features of the deterministic phase-locking pattern. Adding stochasticity to this picture, we have used the Freidlin-Wentzell theory to characterize how the sharp deterministic boundaries of tongues are blurred by the noise. Then, we have extended our stochastic analysis in order to compare two exponentially small scales: the slow-fast scale, estimated by the WKB method and the small noise scale, estimated by LDP. To demonstrate the effectiveness of our scale comparison, we have analyzed the persistence of canard trajectories under the noise.

Our analysis illustrates the quite expected fact: the phase-locking domains are recognizable even with the noise, whereas the canard trajectories have poor observability, since they are the most sensitive trajectories. But even this poor observability can be controlled via the interplay between two exponentially small scales: the WKB method allows us to control deterministic canards, while the Freidlin-Wentzell action provides the estimation of noise effects. Of course, the WKB method is not a panacea for slow-fast dynamics, since we actively exploit the correspondence between M\"{o}bius systems and spectral properties of related Shr\"{o}dinger equations, which is known~\cite{johnson1982rotation,avila2012gaps}.  As for canards studies, our construction complements works related to stochastic selection~\cite{sowers2008canards} and folded-node early-exit theory~\cite{berglund2012hunting}. The WKB matching provides the incoming mismatch and repelling itinerary.

In general situation, merging of fold points and singularities of the Schr\"{o}dinger equation potentials should be analyzed very carefully, since even in the leading order they can affect bandgap structures significantly. Furthermore, when we deal with $g>1$ WKB curve, interference can arise among competing cycle combinations and relative paths, so connection coefficients become decisive. Stochastic analysis can become complicated too. For example, related studies of periodically modulated escape show how driving changes optimal paths and prefactors~\cite{dykman2005activated}. The comparison of two exponentially small scales differs from asking whether escape is likely during observation, which depends on the integrated escape rate. Equal exponents can coexist with rare events. Slower forcing improves adiabatic resolution, but lengthens noise exposure and amplifies fluctuations along a fixed repelling itinerary. Therefore, weak-noise, slow-forcing, and long-observation limits require an explicit ordering or joint scaling.

Due to the exponential sensitivity, canards are hard to detect in experiments, and even in numerical simulations, we need a significant accuracy. Noise makes the observation even more difficult. From a statistical mechanics point of view, the correspondence between the Poincar\'{e} rotation number and the transport current suggests that canards can be associated with rare transport events, but it looks like a vain hope. A much more promising way to observe canards (at least indirectly) is their relation to the so-called \emph{delay of stability loss}.

Delay of stability loss provides a promising extension~\cite{berglund2000dynamic}. Under the slow parameter variation, trajectories can remain near a persisting equilibrium branch after its frozen-system stability changes, a mechanism distinct from passage through a fold bottleneck. For a class of analytic Hopf passages, complex singularities and Stokes geometry determine the maximal deterministic delay~\cite{neishtadt2026maximal} and this suggests a connection with the exact WKB. Noise continually seeds perturbations that are amplified during unstable evolution, as quantified for dynamic pitchfork bifurcations~\cite{berglund2002pathwise}. Our result supplies a concrete analog for prescribed M\"{o}bius canard itineraries. Under the stated entry and covariance assumptions, $D/D_{\text{path}}\to 0$ preserves a fixed slow-time repelling residence with high probability, whereas $D/D_{\text{path}}\to\infty$ removes this passage within the local reduction. It looks that a natural next step is to determine when departure-time distributions retain a measurable signature of exponentially small deterministic connection data despite this stochastic amplification. In periodically forced systems, departure-phase distributions offer a corresponding observable even when the mean winding changes little.

The overdamped Josephson junction also looks prominent as a possible experimental setup. The Shapiro voltage steps~\cite{shapiro1963josephson}, which represent the mean voltage, coincide with the rotation number up to a numerical factor. The completed phase-slip counts and the voltage noise spectra are hidden in this staircase structure. The quasipotential methods~\cite{kautz1994quasipotential} already quantify the phase-locking persistence and noise tolerance. Our framework compares this tolerance with the fine deterministic structures and excursions over unstable branch, suggesting distinct parameter marginis to observe a voltage plateau and to detect canard. We highlight that quantum phase-slips and tunneling phenomena should not be associated with the discussed phase-slips, since the equation Schr\"{o}dinger here is an auxiliary representation of dissipative dynamics.

For the $\theta$-neuron and its QIF representation, the excursion over unstable branch delays motion near a moving firing threshold. Noise can change the firing phase and the spike count per drive period. Relevant measurements include firing-phase distributions, extra or missed spikes, and survival of a threshold-following interval. The $\theta$-QIF transformation also changes the noise coupling. The paper~\cite{franzen2023theta}, related to frequency-dependent firing responses with correlated current noise, suggests an extension requiring the appropriate coupling and color-noise action. In addition, superconducting neuron models and coupled Josephson-neuron experiments offer a neuromorphic connection~\cite{crotty2010josephson,segall2017synchronization}, conditional on a valid scalar reduction and calibrated noise. Capacitance, coupling, and delays can introduce additional dynamics.

Coupled phase oscillators can exhibit common-noise synchronization, which means convergence from different initial conditions under the same realization~\cite{braun2012global}. Since identical phase oscillators with shared drift and noise can be described via M\"{o}bius dynamics, we assume that canard trajectories can arise in such systems in similarity to~\cite{touboul2015noise}, but a separate study is needed.

Finally, spatially deformed conformal field theories provide a further connection through driven products of M\"{o}bius transformations~\cite{wen2021driven}. For suitable random protocols, matrix Lyapunov growth controls heating, with exceptional laws possible when Furstenberg's hypotheses fail~\cite{wen2022furstenberg}. Whether slow stochastic protocols admit WKB estimates for finite-time growth and rare fluctuations remains open and requires an explicit map to quantum observables. The present canard and phase-slip estimates do not establish a heating theorem.

\section*{Acknowledgements}

I am grateful to Andrey Dymov, Mauro Mariani, and Alexey Glutsuyk for the discussions and to Jean Pierre Ramis for sharing Pavis d'Escurac's thesis. This work was performed at the Steklov International Mathematical Center and supported by the Ministry of Science and Higher Education of the Russian Federation (agreement no. 075-15-2025-303). I also acknowledge the support by the Foundation for the Advancement of Theoretical Physics and Mathematics ``BASIS'' (grant No. 25-1-4-2-1). All numerical simulations were performed with {\julia} programming language and library \verb|DifferentialEquations|.

\begingroup
\footnotesize
\bibliographystyle{unsrt}
\bibliography{references}
\endgroup

\end{document}